\documentclass[aps,prb,reprint,showpacs,superscriptaddress]{revtex4-1}

\usepackage{graphics}
\usepackage{bm}
\begin{document}
\preprint{In preparation to be submitted J. Phys Condensed Matter}
\title{Strain driven monoclinic distortion of ultrathin CoO films in CoO/Pt(001) and exchange-coupled CoO/PtFe/Pt(001) systems }
\author{Anne D. Lamirand}
\affiliation{Universit\'e Grenoble Alpes, Institut N\'eel, F-38042 Grenoble, France}
\affiliation{CNRS, Institut N\'eel, F-38042 Grenoble, France}
\author{M\'arcio M. Soares}
\affiliation{European Synchrotron Radiation Facility - ESRF, Grenoble, France}
\author{Maurizio De Santis}
\email[]{maurizio.de-santis@neel.cnrs.fr}
\affiliation{Universit\'e Grenoble Alpes, Institut N\'eel, F-38042 Grenoble, France}
\affiliation{CNRS, Institut N\'eel, F-38042 Grenoble, France}
\author{Aline Y. Ramos}
\email[]{aline.ramos@neel.cnrs.fr}
\affiliation{Universit\'e Grenoble Alpes, Institut N\'eel, F-38042 Grenoble, France}
\affiliation{CNRS, Institut N\'eel, F-38042 Grenoble, France}
\author{St\'ephane Grenier}
\affiliation{Universit\'e Grenoble Alpes, Institut N\'eel, F-38042 Grenoble, France}
\affiliation{CNRS, Institut N\'eel, F-38042 Grenoble, France}
\author{H\'elio C. N. Tolentino}
\affiliation{Universit\'e Grenoble Alpes, Institut N\'eel, F-38042 Grenoble, France}
\affiliation{CNRS, Institut N\'eel, F-38042 Grenoble, France}
\date{\today}

\begin{abstract}
The structure and strain of ultrathin CoO films grown on a  Pt(001) substrate and on a ferromagnetic  PtFe pseudomorphic layer on Pt(001) have been determined with $\it{in} {situ}$ and real time surface x-ray diffraction. The films grow epitaxially on both surfaces with an in-plane hexagonal pattern that yields a pseudo-cubic CoO(111) surface. A refined x-ray diffraction analysis reveals a slight monoclinic distortion at RT induced by the anisotropic stress at the interface. The tetragonal contribution to the distortion results in a ratio $\frac{c'}{a'}>1$, opposite to that found in the low temperature bulk CoO phase. This distortion leads to a stable Co${^{2+}}$ spin configuration within the plane of the film. 
\end{abstract}

\maketitle
\section{\label{intro} introduction}
 
Critical phenomena at the nanoscale size follow a different behaviour than in bulk. A reduction of critical temperature with thickness is expected in magnetic films when their size is comparable to the characteristic correlation length associated to the interatomic magnetic interaction. This intrinsic effect is often hindered by unwanted changes in the structure and in the chemical composition, as well as by interface effects. 
The importance of interface parameters tends to increase with the constant seek for reduction of dimensions in technological systems. The complete structure of these systems is a key to properly disentangle finite size, bulk and surface structural effects.

Due to its technological applications the exchange bias (EB) effect, which takes place at ferromagnetic-antiferromagnetic (FM-AFM) interfaces, has been widely investigated in the last decades. Below a critical  blocking temperature  (${T}_{B}$), the coupling between the two layers induces a shift of the hysteresis loop of the FM layer due to an intrinsic exchange bias field ($H_{EB}$) \cite{Meiklejohn1956PR, Nogues1999}. Despite its scientific relevance and its widespread use in magnetoelectronic applications, several aspects of this effect have not been completely elucidated, especially the assessment of the magnetic  structure at the interface and throughout the volume of the AFM layer\cite{Morales_PRB09, Chappert2007NMat}. This latter is strongly influenced by the interface structural parameters \cite{Kiwi_JMMM2001} as reduced coordination of surface atoms, atomic modifications induced by strain or chemical diffusion at interfaces. For example, an apparent discrepancy between the amplitude of the reduction of the N\'eel temperature (${T}_{N}$) with the thickness in nano-periodic CoO/SiO${_{2}}$ \cite{Ambrose_1996} and CoO/MgO multilayers \cite{Abarra_1996} was explained several years later by the presence of an amorphous CoO layer at the interface with SiO${_{2}}$. \cite{Tang_2003}

Bulk CoO is AFM with ${T}_{N}$  of 293 K and a large magnetic anisotropy. In the paramagnetic phase, it crystallizes in rock-salt structure with ${a_{CoO}}$=4.261 {\AA}. Co and O planes alternate along the [111] directions, with an in-plane atomic arrangement given by an hexagonal mesh and an interatomic distance of 3.013 {\AA}. The AFM transition goes along a cubic-to-monoclinic crystallographic distortion. At 10 K, the monoclinic constants are a=5.18190(6) {\AA}, b=3.01761(3) {\AA}, c=3.01860(3) {\AA} and  ${\beta}$=125.5792(9), with  ${\beta}$ the angle between ${a}$ and ${c}$.\cite{Jauch2001PRB} In the pseudo-cubic face-centered setting, this corresponds to an angle of 89.962 between the two edges of different lengths. Its magnetic structure is collinear with high spin Co${^{2+}}$  ions forming a stacking of FM planes coupled antiferromagnetically along the pseudo-cubic [111] axis. The moments are oriented in the monoclinic ${ac}$ plane, pointing close to the  [001] axis\cite{Jauch2001PRB, Roth1958PR}. This magnetic structure, usually denoted as AFM II \cite{Roth1958PR}, is stabilized by superexchange mediated by the oxygen atoms\cite{Anderson_1950, Anderson_1963}.

In films with nanometric thicknesses, the CoO structure is extremely influenced by the substrate, through the epitaxial strain and interface chemistry, as well as by the growth conditions. CoO grows epitaxially on Ag(001) with the same (001) orientation and bulk-like structure, except for a slight in-plane compressive strain. It grows pseudomorphic for a thickness of 4 ML \cite{Schindler2009} but then it relaxes with the appearence of misfit dislocations. On Pd(001), Gragnaniello et al.\cite{Gragnaniello2010a} reported a modification of the CoO cristallographic orientation and morphology as function of the coverage. Meyer et al. showed the presence of a wurtzite-like CoO bilayer at the surface of CoO(111)/Ir(100) films\cite{Meyer_PRL08}. Moreover, on this substrate the growth orientation can be switched from (111) to (100) by the predeposition of a 1ML Co buffer layer \cite{Gubo_PRL12}. On Pt(111), CoO films are (111) oriented, and a Moir\'e pattern can be observed according to the oxidation conditions and thickness of the deposited layer \cite{DeSantis2011PRB}. These structural differences are likely to strongly influence the magnetic properties of the films, as nicely demonstrated by a spin reorientation related to the nature -compressive or expansive- of the strain \cite{Csiszar2005PRL}. Unexpectedly, the growth of CoO on Pt(001) has never been reported in literature. Moreover, epitaxial CoO film on Pt(001) opens the way towards a CoO/FePt AFM/FM bilayer on Pt, since Pt-terminated FePt(001) and Pt(001) present rather the same chemical surface.

In this paper we report epitaxial growth of CoO films on a Pt(001) substrate and on a FM PtFe(001) pseudomorphic alloy on Pt(001). We have succeeded at the growth of strained ultrathin (111)-oriented CoO films on both Pt(001) and Pt-terminated PtFe(001) surfaces and have precisely determined the strain and the structure of CoO layers with $\it{in} {situ}$ and real time surface x-ray diffraction. We find that on both surfaces the CoO structure presents a slight monoclinic distortion stable at room temperature (RT) evocative of the low temperature and magnetically driven distortion of the bulk. At local scale, such a distortion translates into a tetragonal distortion of the octahedra formed by the first oxygen neighbors around Co atoms. The implications of that distortion on magnetism and spin orientation of Co${^{2+}}$  ions is discussed.

\section{\label{experim}experimental}

Surface x-ray diffraction experiments at grazing incidence (SXRD) were performed at the French CRG-IF BM32 beamline \cite{BaudoingSavois_1999_213} at the European Synchrotron Radiation Facility (ESRF) using a photon energy of 22 keV (${\lambda}$=0.56354{\AA}). The ultrahigh vacuum (UHV) chamber, fully equipped for sample preparation, was working at a base pressure of ${P }$${\sim 2\times 10^{-10}}$ mbar. The Pt(001) single-crystal substrate was prepared by cycles of 800 eV ${Ar^+}$ sputtering and annealing at about 1200 K. After preparation no impurities were detected by Auger electron spectroscopy and SXRD showed large terraces of the typical quasi-hexagonal reconstruction \cite{AbernathyDL_PRB1992, Soares2012PRB}.  Fe, Co and Pt were deposited by molecular beam epitaxy (MBE), using pure rods inserted in water-cooled electron beam evaporators. The deposition rate, calibrated with a quartz crystal microbalance, was typically 1 monolayer (ML) per 10 minutes for Fe and Co, and 1 ML per 45 minutes for Pt. CoO films were grown with  the Pt substrate held at 520 K by reactive deposition of Co, keeping a molecular oxygen pressure in the chamber of  ${P_{O_{2}}}$${\sim 5\times 10^{-7}}$ mbar during evaporation.

The diffracted intensity as function of momentum transfer (${Q=\frac{4 \pi sin \theta}{\lambda}}$) from a truncated crystal (or an ultrathin epitaxial film) shows sharp scattering line-shape parallel to the surface and continuous line-shape in the out-of-plane direction. These out-of-plane intensity distributions at fixed  (HK) indexes  of the surface mesh are known as (HK) difrraction rods. When originating from the substrate they are called crystal truncation rod (CTR) \cite{RobinsonIK_PRB1986}. The surface unit-cell is defined by the three vectors  ${\vec{a}_{Pt,i}}$ of a tetragonal body centered mesh related to the fcc substrate by ${\vec{a}_{Pt,1}=\frac{a_{Pt}}{2}\text{ⅹ}[1\overline{1}0]}$, ${\vec{a}_{Pt,2}=\frac{a_{Pt}}{2}\text{ⅹ}[110]}$, and ${\vec{a}_{Pt,3}=a_{Pt}\text{ⅹ}[001]}$  with ${a_{Pt} = 3.924}$ \AA. An (HK) CTR bind bulk Bragg peaks with the same (HK). When a film grows in registry with the substrate, film and substrate contributions interfere on the same rod. The film structure has to be resolved by a fine analysis of these CTRs intensities \cite{VliegE_JACrys2000:ROD}. It requires quantitative data, which are obtained by an integration of the scattered intensity at (HKL) nodes along the CTR, by rocking the sample around the surface normal (CTR-scan). Then the structure factors are extracted applying standard correction factors \cite{VliegE_JACryst1997} and averaged over symmetry equivalent reflections.  The data reduction and structural analysis are performed using the ANA-ROD package that uses a ${X^{2}}$ minimization for fitting occupation profiles and interplane distances \cite{VliegE_JACrys2000:ROD, VliegE_JACryst1997}.

After optimizing the growth procedure, two samples were studied in detail by SXRD. The first one is a nominally 3.5-nm-thick CoO layer grown on a clean Pt(001) substrate held at 520 K by reactive (${P_{O_{2}}}$${\sim 5\times 10^{-7}}$ mbar) Co deposition. The second one is a CoO/FePt/Pt(001) double-layer film, where the CoO layer is coupled to a FM L1${_{0}}$  ordered PtFe layer in registry with the Pt(001) substrate. The FePt layer was grown by evaporation of nominally 3 ML of Fe on a clean Pt(001) substrate held at 600 K, followed by 1 ML Pt deposition. This strategy of FePt growth steems from the fact that an atomic site exchange process takes place between Fe adatoms and Pt atoms of the substrate \cite{He_PRB2005}. At higher temperatures, the process is reinforced and the Fe adatoms interdiffuse into the Pt substrate to form an epitaxial surface alloy \cite{Soares2011JAP}. The additional Pt monolayer ensures a Pt rich interface with CoO, intended to limit Fe oxidation and produce a similar interface chemistry as for a clean Pt substrate. The resulting PtFe(001) layer is pseudomorphic with the substrate. No traces of L1${_{2}}$  type ordered alloys or L1${_{0}}$ domains with the c-axis lying in-plane were found. Then, a CoO layer was grown at 520 K starting by deposition of 2 ML of metallic Co followed by oxidation during 10 min and additional reactive (${P_{O_{2}}}$${\sim 5\times 10^{-7}}$ mbar) Co deposition up to a 4-nm-thick CoO layer. 
For the analysis of these non-pseudomorph CoO layers, the intensity of the additional rods appearing between those of the substrate was measured simply by scanning the momentum transfer perpendicular to the surface (L-scans). The structure factors were extracted using the so called stationary geometry corrections, after a background substraction \cite{VliegE_JACryst1997}. The structure of the CoO layer turn out to be essentially the same on both Pt(001) substrate and  Pt-terminated PtFe(001) FM layer. 

To provide a fully structural description of the exchanged-coupled system, a total of 128 non-equivalent reflections belonging to the (10), (11), and (20) CTRs (as indexed in the surface unit-cell) were used for the determination of the atomic structure of the PtFe layer. The PtFe layer rods measured before and after CoO growth show minor changes ascribed to some oxidation of  Fe at the interface. 
In order to estimate the extent of this oxidation, x-ray absorption spectroscopy (XAS) at the Fe ${L{_{2,3}}}$  edges was performed $\it{ex} {situ}$ at the soft x-ray beamline of the ESRF. All spectra were collected with a spectral resolution of ${E/\Delta E}$ ${\sim6000}$ and using total electron yield, corrected for electron yield saturation effects and normalized far from $L_{2,3}$  edges.

\section{\label{results}results and discussion}
\subsection{Structure of PtFe layer and interfaces}

\begin{figure}
\resizebox{1\columnwidth}{!}{\includegraphics{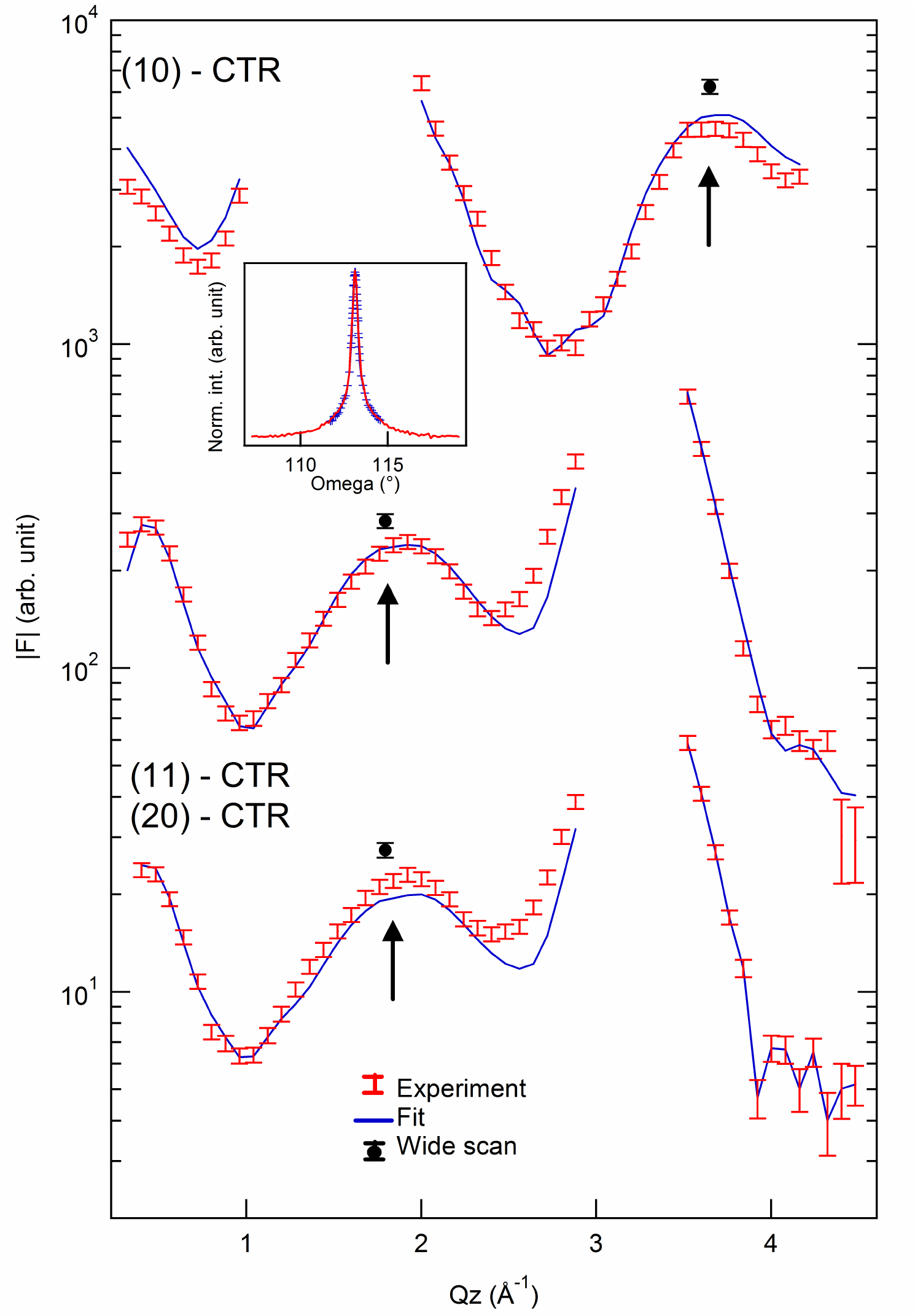}}
\caption{\label{fig:PtFeCTR} (color online) X-ray scattering through the Pt rods ((10), (11) and (20) CTRs) after FePt and CoO growth. The scattered marks (experiment) are associated to error bars calculated from data dispersion of symmetry equivalent rods. Calculated structure factor amplitudes are represented by solid lines. The broad peaks associated to the doubling of the electronic density in PtFe are pointed by arrows. Black circles represent experimental structure factor amplitudes extracted from much wider angular scans (inset) for the integration of the CTR intensity.}
\end{figure}

Reference scans were collected for the clean  substrate prior to metal deposition. They show, besides the CTRs, additional rods coming from the typical Pt(001) quasi-hexagonal reconstruction\cite{AbernathyDL_PRB1992, Soares2012PRB}. After Fe deposition, the reconstruction is lifted and scans of the momentum transfer parallel to the surface (${Q_{//}}$) do not show any extra peak or shoulder close to the CTR. These measurements, repeated at different values of the momentum transfer perpendicular to the surface (${Q_{z}}$), ensures that the layer is in registry with the substrate, without any in-plane relaxation. The CTR structure factors reflect then the interference between the scattering from the FePt film and the Pt substrate. Broad peaks have shown up between the substrate Bragg peaks at ${Q_{z}\sim3.68}$  {\AA}${^{-1}}$  for (1 0) and at ${Q_{z}\sim1.84}$  {\AA}${^{-1}}$ for (1 1) and (2 0) CTRs (pointed by arrows in Fig.\ref{fig:PtFeCTR}), indicating a doubling of the electronic density period along the growing direction. These order peaks correspond to the ${L1_{0}}$ phase with c-axis perpendicular to the surface \cite{Soares2012PRB, Soares2011JAP}. The lattice mismatch favours this orientated growth because the bulk FePt ${L1_{0}}$ lattice parameters, ${a_{bulk}}$=3.860 {\AA} and ${c_{bulk}}$=3.713 {\AA} are smaller than ${a_{Pt}}$   by 1.6\% and 5.4\%, respectively.


\begin{table}
\centering{}%
\begin{tabular}{|c|c|c|c|c|c|}
\hline 
{layer (n)} & \multicolumn{2}{c|}{region} & {$d_{n,n-1}$} & $occ_{Pt}${/}$occ_{Fe}$ & $occ_{Pt}'${/}$occ_{Fe}'$\tabularnewline
\hline 
\hline 
{1} & \multicolumn{2}{c|}{} & {1.97 (1)} & {1/0} & \tabularnewline
\cline{4-5} 
{2} & \multicolumn{2}{c|}{interface with Pt} & {1.99 (1) } &{0.7(1)/0.3} & \tabularnewline
\hline 
{3} & {\small{FePt}
} & {\footnotesize Fe plane} & {{1.804(5) }} & {{} 0.10(5)/0.9} & {{} 0.1/0.9}\tabularnewline
\cline{6-6} 
{4} & {alloy} & {\footnotesize Pt plane} & {{1.804(5) }}  & {{} 0.7(1)/0.3} & {0.8/0.2}\tabularnewline
\cline{1-1} \cline{3-3} \cline{5-6} 
{5} & {film}  & {\footnotesize Fe plane} & {{1.804(5) }}  & {{} 0.2(1)/0.8 } & {0.1/0.9 }\tabularnewline
\cline{6-6} 
{6} &  & {\footnotesize Pt plane} & {{1.804(5) }}  & {0.7(1)/0.3} & {0.8/0.2}\tabularnewline
\hline 
{7} & \multicolumn{1}{c}{} &  & {1.93(2) } & {0.5(1)/ -} & {}\tabularnewline
{8} & \multicolumn{2}{c|}{{interface with CoO}} & {1.80(4) } & {{} 0.4(1)/ -} & \tabularnewline
{9} & \multicolumn{1}{c}{} &  & {1.97(2) } & {0.15(5)/ -} & \tabularnewline
\hline 
\end{tabular}\caption{\label{Table1} Fitting results of a 4 ML FePt alloy
and its interfaces after cobalt oxide deposition. Occupation rates
($occ_{Fe}$ and $occ_{Pt}$) and interplane distance $d_{n,n-1}$
to the preceeding layer are described for each layers. The last column
($occ'_{Fe}$ / $occ'_{Pt}$) reports an estimation of the occupancies
obtained {thanks to wide angular scans performed
at the superstructure peaks maximum.}}
\end{table}

As stated here above, the FePt film grows in registry with the substrate. The layer adopts the in-plane  lattice parameter of Pt. The film structure parameters are optimized through a best fit of the calculated structure factors to the measured ones \cite{VliegE_JACrys2000:ROD}. The model consists of a partially ordered FePt(001) film between two interface regions: one with the Pt substrate and the other with the CoO overlayer. For each layer ${n}$, atoms occupy bulk-like sites and the fit parameters are occupancy rates (${occ_{Fe}}$  and ${occ_{Pt}}$) and interplane distance ${d_{n,n-1}}$  to the previous layer, with ${occ_{Fe}+occ_{Pt}=1}$ for each layer (except at the CoO interface). An average Debye-Waller has been considered for the whole alloy layer. The Pt interface is defined by two atomic layers. Layer 1 contains Pt only, while Fe is allowed to diffuse to the layer 2. ${d_{1,0}}$ and ${d_{2,1}}$ distances are free and independent. Layers 3 to 6 correspond to the FePt alloy film, with a single interlayer distance ${d_{alloy}}$.  Fe and Pt occupation rates were fitted on each layer. For the CoO interface region ${(n>6)}$, distances were free and independent. Fe and Pt occupation rates were adjusted with the only constraint of ${occ_{Fe}+occ_{Pt}<1}$. 
The best fit Debye-Waller factor, ${B_{alloy}=0.96}$, turn out to be larger than the Fe and Pt bulk ones (0.39 and 0.31, respectively) indicating that the structural disorder dominates and justifying the use of a single parameter. The best fit parameters are summarized in Table \ref{Table1} and the best fit curves are plotted in Fig.\ref{fig:PtFeCTR} along with data.

 Some outcomes should be emphasized. In the fully ordered ${L1_{0}}$ phase, pure Pt and Fe planes alternate along the ${c}$-axis. The occupancies obtained for the FePt alloy from ${n=}$3 to 6 (Table \ref{Table1}) correspond to alternate Fe rich and Pt rich layers. The degree of order can be quantified by the parameter ${S}$, defined in a binary alloy as ${S=r_{a}-w_{a}}$, where ${r_{a}}$  and ${w_{a}}$  are the ratio of ${a}$-sites occupied by the right and by the wrong kind on atoms, respectively \cite{Warren_1969}. ${S=0}$ (${S=1}$) for a complete disordered (perfectly ordered) alloy. Applying this definition to the layers from 3 to 6, we have ${S=0.5(1)}$ on average. Actually, this value is largely underestimated because the intensity coming from small ordered domains is scattered over a large reciprocal space region and is not fully integrated during the standard rocking scans over the CTRs. A more accurate order parameter is obtained by performing wider angular scans (inset Fig.\ref{fig:PtFeCTR}) about the order peaks (indicated by arrows in fig.\ref{fig:PtFeCTR}). The corresponding reevaluated structure factors are reported in Fig.\ref{fig:PtFeCTR}. Within the same model, the Pt and Fe occupancies for layers 3 to 6 were adjusted to match the corrected structure factors, all other structural parameters being unchanged. This results in the final Fe and Pt occupancies reported in the last column of Table \ref{Table1}. One ends up with an order parameter of ${S\sim0.7(1)}$. 

The lattice constant ${c_{alloy}=2d_{alloy}}$=3.61(1) {\AA} is compressed by 2.8\% with respect to bulk FePt one. This compensates the (001) plane atomic density reduction, which results from the substrate induced in-plane strain. The tetragonal distortion of this pseudomorphic layer ${(c/a)_{alloy}=0.920(3)}$ is enhanced compared to the bulk value ${(c/a)_{bulk}=0.962}$. A similar value was previously measured on a 2.0-nm-thick FePt film grown by alternate deposition on Pt(001) \cite{Soares2011JAP}. 

The interface with the CoO is a mixture of different phases and its complete structure can be hardly obtained by SXRD. It is then simply modeled with partially occupied layers of lonely Pt. It should be pointed out that in these layers a Pt fraction could be replaced by a larger Fe fraction, which has a lower atomic scattering factor, whithout a relevant increase of the ${X^{2}}$. Therefore, the reported Pt occupancies at the interface represent an upper limit. It should be kept in mind that, if some amount of Fe is present at that interface,  it has to be Pt covered, otherwise it would form a relaxed oxide that does not contribute to CTRs. To wrap up, the PtFe interface with the CoO layer is satisfactorily described by a mixture of Pt covered metallic Fe and Pt islands embedded in a relaxed Co (Fe) oxide layer. Coexistence of CoO and Pt islands at the surface of a CoPt alloy has already been reported in literature \cite{DeSantis2011PRB}.

\begin{figure}
\resizebox{1\columnwidth}{!}{\includegraphics{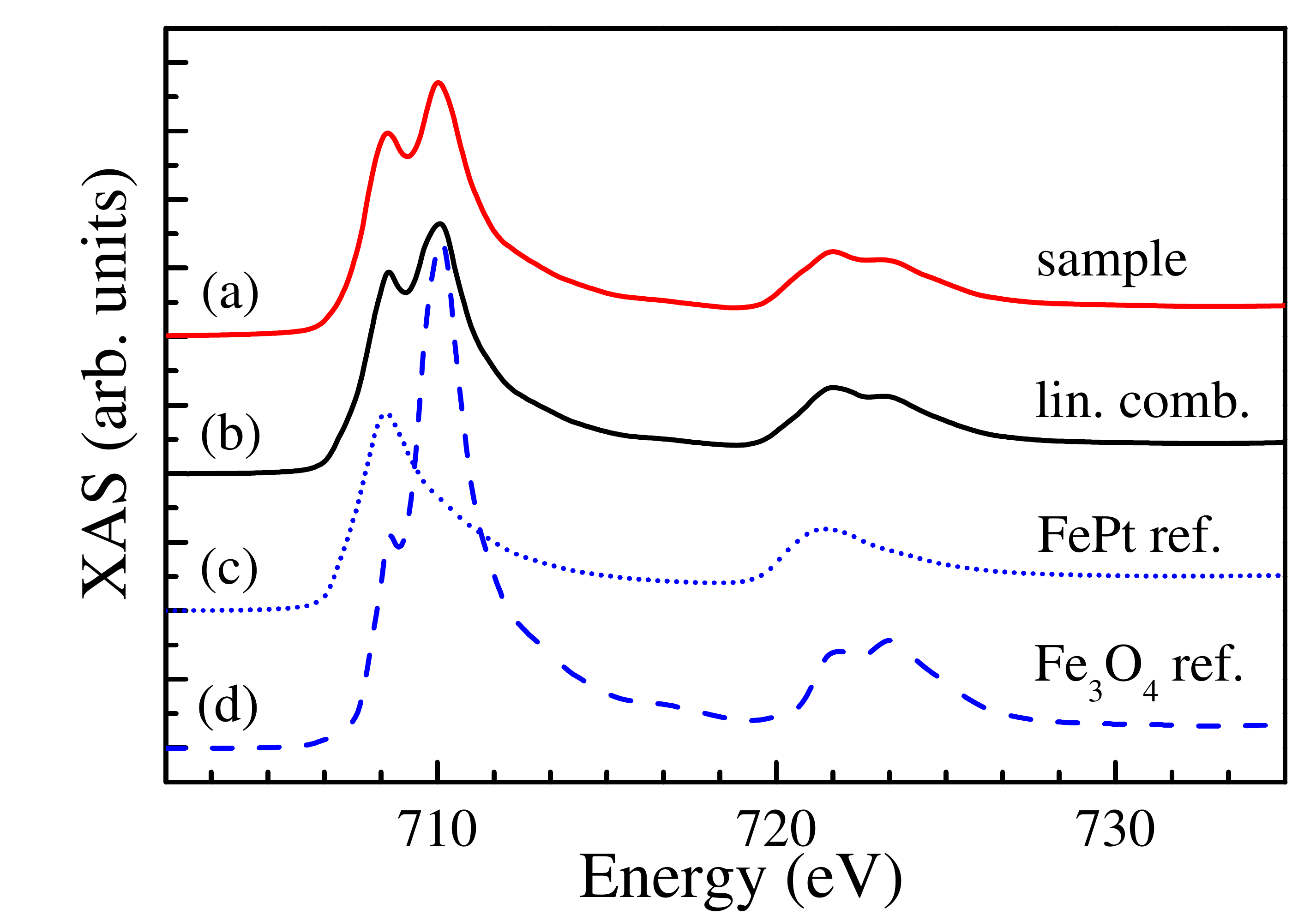}}
\caption{\label{fig:XASFeL} (color online) Comparison of CoO/PtFe/Pt(001) sample
${L_{2,3}}$-edges XAS (a) with a linear combination (b) of two reference spectra: metallic
FePt (c) and Fe$_{3}$O$_{4}$/CoO (d).}
\end{figure}

The total Fe content summed over the metallic alloy film (n=1 to 6)  is 2.6 ML, smaller than the nominal 3 ML value, highlighting a partial Fe oxidation. The extent of this oxidation is estimated by comparing the XAS spectrum of the CoO/FePt/Pt(001) sample with those of a metallic FePt and a double-layer CoO(4nm)/Fe${_3}$O${_4}$(2nm)/Ag(001),  previously prepared and used as references. Figure \ref{fig:XASFeL} shows a linear combination (full line) of the two reference spectra with a weight 0.35 for the Fe oxide (dash) and 0.65 for the metallic Fe (dot). It gives a very good agreement over the entire spectrum, including the multiplet structure at the ${L_{2}}$ edge. Taking 2.6 ML (65\%) as the metallic contribution, we found out that approximately 1.4 ML (35\%) of Fe are an oxide environment.

Fe oxidation close to the interface is expected owing to its high oxidation potential \cite{Regan2001PRB}. Reactive CoO deposition on pure Fe results in the oxidation of at least 2 ML of Fe at 340 K \cite{Bali_APL12}. Our Pt-terminated PtFe(001) layer shows a smaller oxide contribution even if the deposition temperature (520 K) was higher than that. Such small oxide contribution is likely related to Fe atoms dispersed within the CoO layer or from Fe-O bounds at the interface. It demonstrates the good resistance to corrosion of the PtFe alloy film. As a matter of fact, the Fe XMCD spectrum of this sample is characteristic of metallic ${L1_{0}}$ PtFe and has no noticeably contribution from the Fe oxide content  \cite{Lamirand_PRB13}. Diffusion of Co atoms into the metallic PtFe alloy prior to the oxidation is discarded, since no metallic component was observable in the XAS measurements around the Co ${L_{2,3}}$  edges \cite{Lamirand_PRB13}.

\subsection{Structure of the CoO layer}

\begin{figure}
\resizebox{1\columnwidth}{!}{\includegraphics{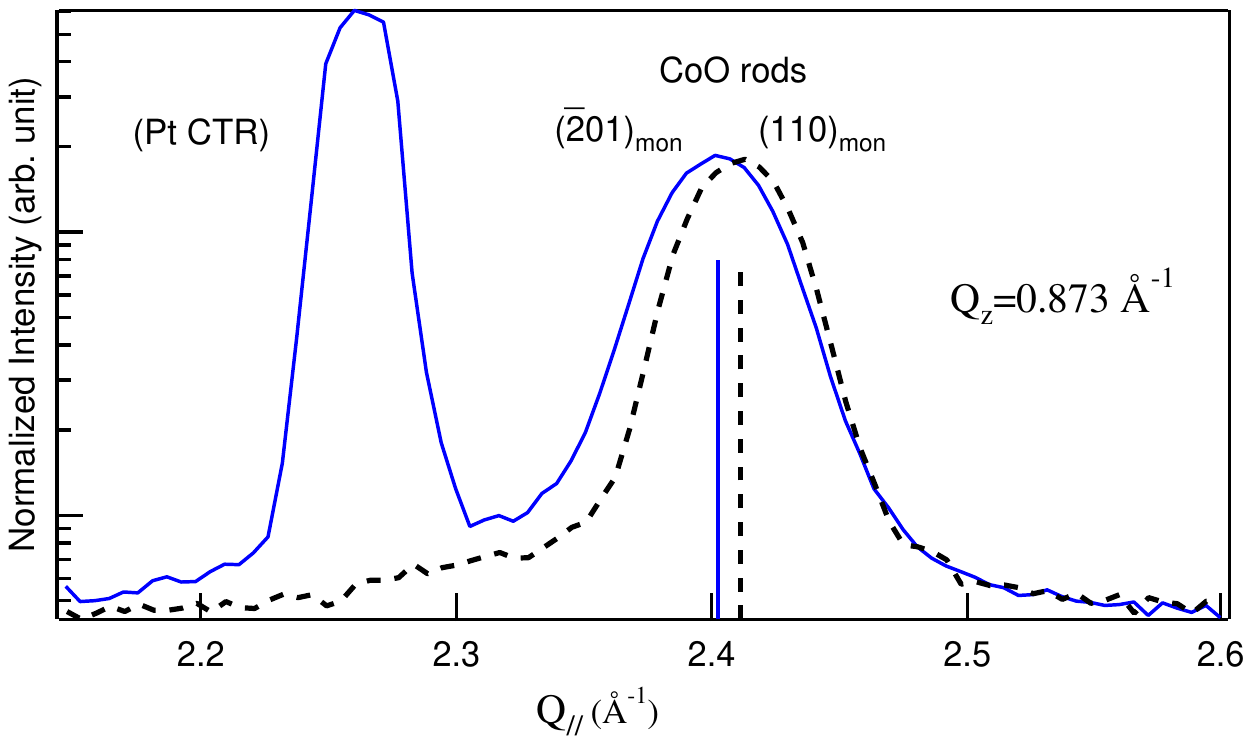}}
\caption{\label{fig:Inplane2} (color online) Intensity distribution as function of the momentum transfer modulus parallel ($Q_{//}$) to the surface at $Q_{z}$=0.873 $\mathring{A}^{-1}$. The plain (blue) line corresponds to a scan along $\vec{a}_{Pt,1}$ and crosses both Pt and CoO rods. The dashed (black) line corresponds to a scan along a direction rotated by 60$^{\circ}$ with respect to $\vec{a}_{Pt,1}$. The shift between the CoO rod positions, pointed by the plain (blue) and dashed (black) lines, reveals the distortion of the in-plane hexagonal pattern.}
\end{figure}

\begin{figure}
\resizebox{1\columnwidth}{!}{\includegraphics{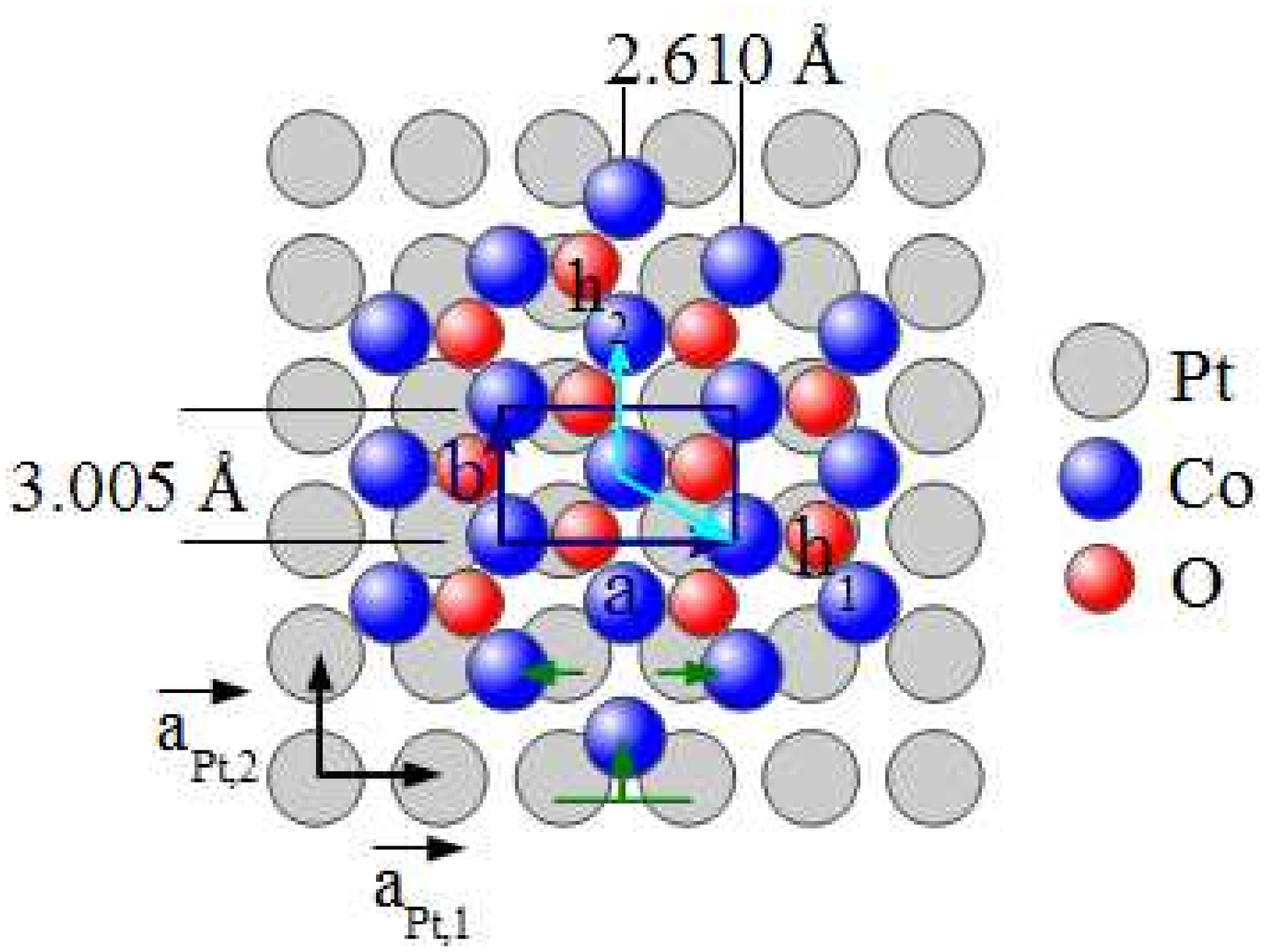}}
\caption{\label{fig:DirectSpace} (color online) Top view of the CoO structure on Pt(001) (or PtFe(001)). The non-regular hexagon is described by a rectangular surface mesh formed by the in-plane vectors $\vec{a}$ and $\vec{b}$ .}
\end{figure}

After CoO growth, scans of the momentum transfer parallel to the surface (${Q_{//}}$) have been performed along different directions in the reciprocal space. They show that the CoO diffraction pattern exhibits a sixfold in plane symmetry, like a (111)-oriented film with twinned domains. Comparing two scans performed along ${\vec{a}_{Pt,1}}$  and in the direction rotated by 60$^{\circ}$  in-plane, respectively, a small but significant shift of the rod position is clearly detected  (Fig.\ref{fig:Inplane2}), indicating that the hexagonal mesh is non regular.  From the rod FWHM we estimate a characteristic domain size of about 8 nm parallel to the surface. By successive rotation of the sample by 90$^{\circ}$ around the surface normal, four growth variants are observed. They are due to the fourfold symmetry of the substrate.  An in-plane schematic view of one of these variants is given in Fig.\ref{fig:DirectSpace}. For the sake of clarity only the two variants related by mirror symmetry are discussed in this paper, but reflections of all variants have been collected, treated and averaged to improve data's quality. The rod position and shape and the related strains are qualitatively the same for the CoO film grown on the clean substrate and on the Pt-terminated FePt/Pt(001) surface, highlighting the same structural behavior. The CoO parameters will be given in the following for the last sample, which is more interesting for its magnetic behaviour and for which a larger data set is available. 

The in-plane lattice constants of the non-regular CoO hexagonal mesh ${h_2}$ (parallel to ${\vec{a}_{Pt,2}}$) and ${h_1}$ (at 120$^{\circ}$) are evaluated from the rod positions. Their values are ${h_2}$=3.005(1) {\AA} and ${h_1}$=3.012(1) {\AA}. While this second value is close to the bulk one for RT CoO rock-salt structure (${\frac{a_{CoO}}{\sqrt{2}}}$=3.013 {\AA}) the first one is clearly below, revealing an in-plane contraction of the CoO structure. The CoO epitaxy is characterized by a misfit between the unit-mesh length ${h_2}$ and the substrate surface row spacing ${a_{Pt,2}}$ (${\frac{h_2-a_{Pt,2}}{a_{Pt,2}}}$) of about +8\%, and between the CoO row spacing ${\frac{1}{2}}$${\sqrt{{(2\times h_1)}^2-h_2^2}}$ = 2.610 {\AA}  and ${a_{Pt,1}}$  of -6\%. This anisotropic stress is responsible for the observed strain.
Green arrows in Fig.\ref{fig:DirectSpace} indicate the sense of displacement of the atoms of Co related to the regular hexagon. As we will see later, the oxygen atoms are also displaced, in such a way that the hexagonal stacking is slightly inclined related to the normal. 

\begin{figure}
\resizebox{1\columnwidth}{!}{\includegraphics{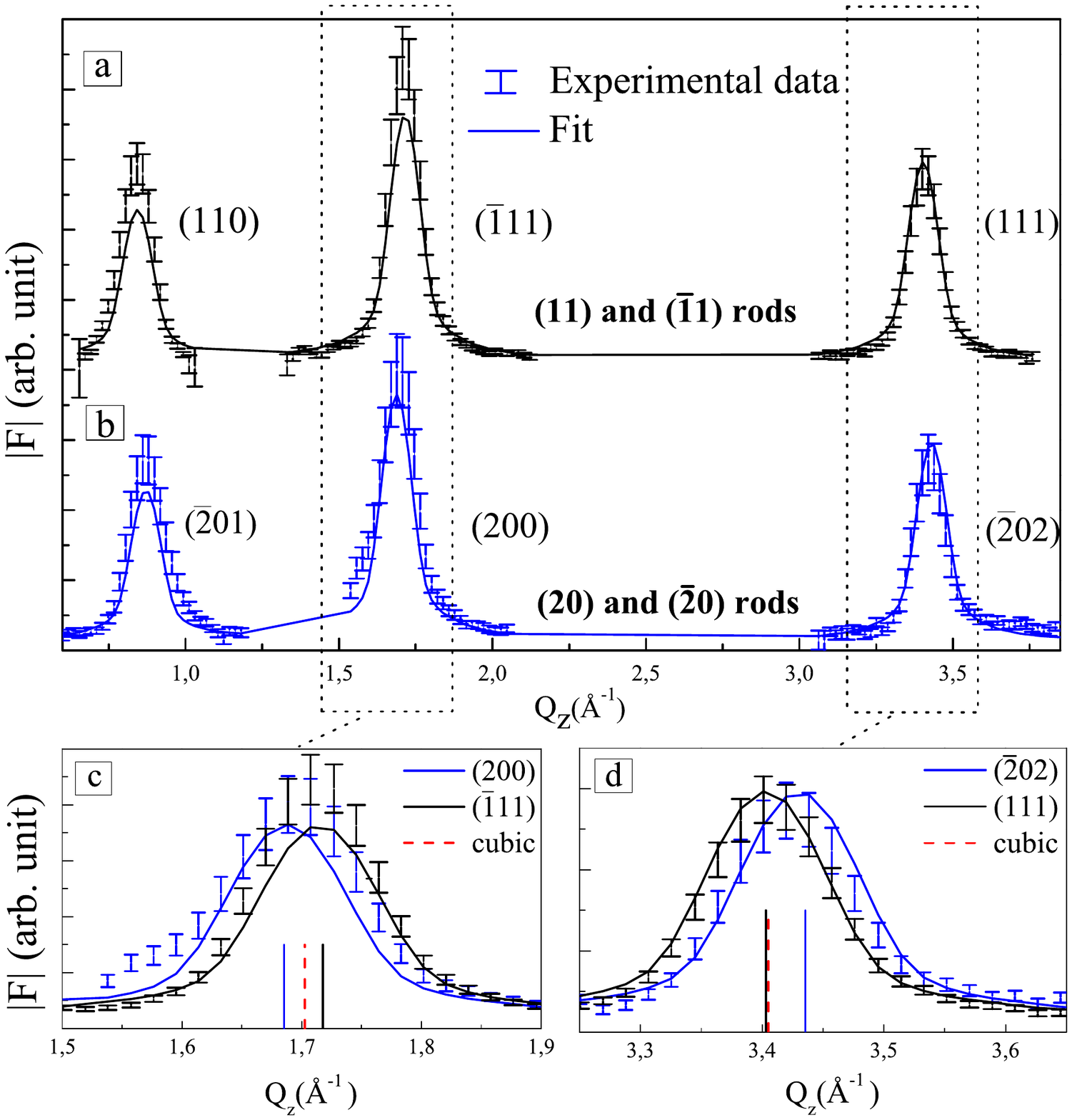}}

\caption{\label{fig:CTR_5} (color online) X-ray scattering through the CoO rods: $(\bar{1}1)$ plus $(11)$ (a) and
$(\bar{2}0)$ plus $(20)$ (b); enlarged view of (200) plus $(\bar{1}11)$ (c) and of $(\bar{2}02)$ plus $(111)$ Bragg peaks.  Scattered marks correspond to experimental data and plain lines are obtained by the structural refinement procedure. Vertical lines indicate the two peaks in monoclinic (plain blue) and in cubic (dashed red) cells. Peak positions for the cubic cell are calculated using CoO bulk data from \cite{Jauch2001PRB}.
 }
\end{figure}

\begin{figure}
\resizebox{1\columnwidth}{!}{\includegraphics{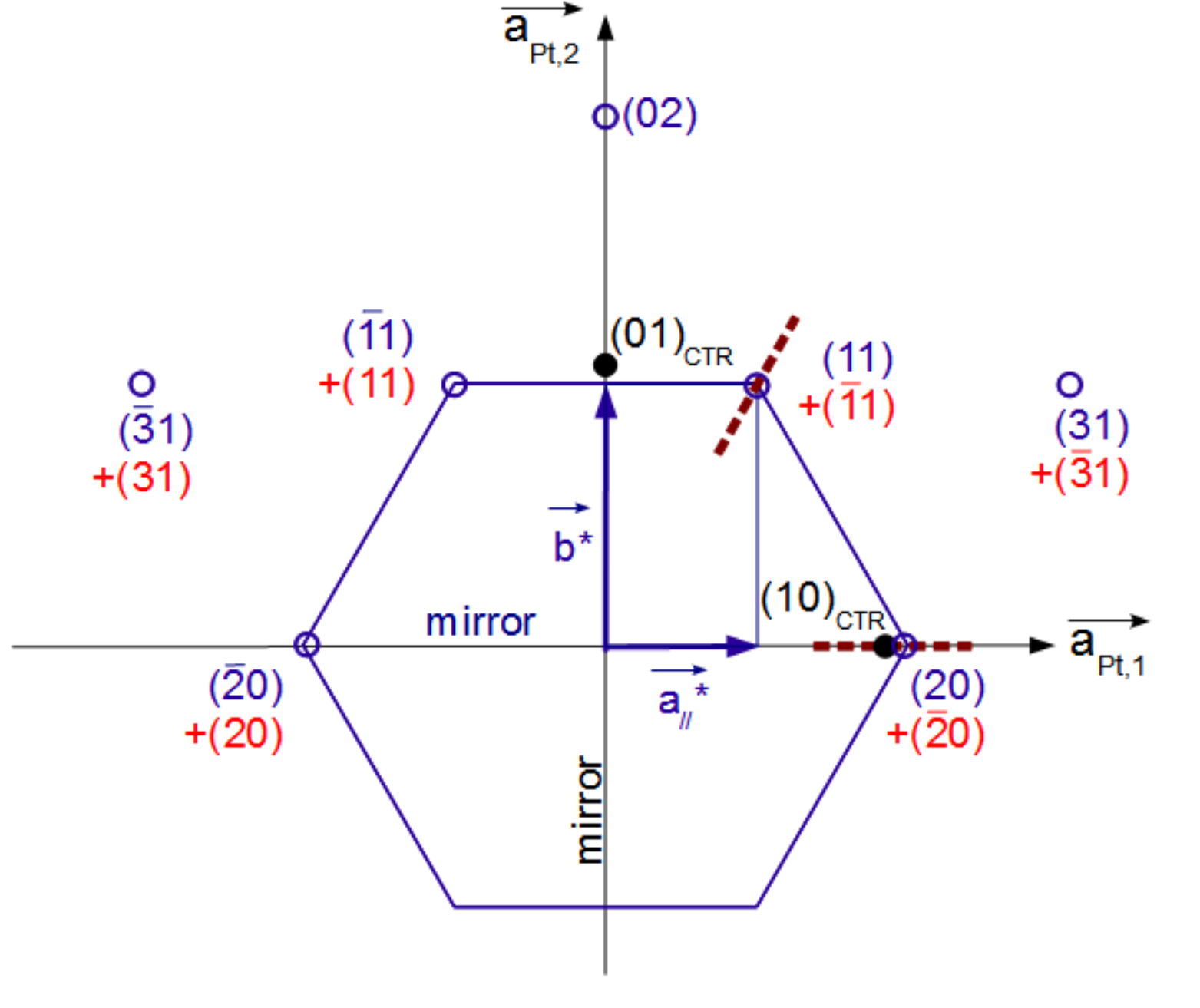}}
\caption{\label{fig_diffraction pattern variants}(color online) In-plane rod positions for the CoO film (blue open circles) and Pt substrate (black plain circles). The CoO diffraction pattern exhibits a quasi-hexagonal symmetry parallel to the surface plane. By inversion symmetry, $(hk)$ rods of one variant (indexed in red) merge with the $(\overline{h}k)$ rods of the mirror variant. }
\end{figure}

The analysis of the structure of the CoO film is founded on the in-plane position of several rods and their intensity
distribution versus the momentum transfer perpendicular to the surface $Q_{z}$ (Fig.\ref{fig:CTR_5}). It takes into account the four growth variants due to the substrate symmetry and requires a clear identification of the diffraction features associated to each of them. If we use a rectangular surface mesh, defined by the vector $\vec{a}$ parallel to $\vec{a}_{Pt,1}$ and $\vec{b}$ parallel to $\vec{a}_{Pt,2}$ (Fig.\ref{fig:DirectSpace}), the four variants correspond to the rotation of this unit cell by the fourfold symmetry axis normal to the substrate. For each variant the positions of the CoO diffraction rods form a non-regular hexagonal mesh on the surface plane. These positions are represented in Fig.\ref{fig_diffraction pattern variants} for two variants related to each other by a rotation of 180$^{\circ}$. These variants can be seen as mirror variants and their relation can also be expressed as an inversion of the coordinate along $\vec{a}$. In the reciprocal space $(hk)$, rods of one variant merge with the $(\overline{h}k)$ rods of its mirror variant. The second set of mirror variants is deduced by 90$^{\circ}$ rotation of this set. Their rods are clearly separated from those of the first set and are not represented.

For the CoO(111) film with cubic rock-salt phase, the rods labeled here (20) and $(\bar{1}1)$ are equivalent, the surface normal being a three-fold symmetry axis. These rods merge with the $(\bar{2}0)$ and (11) ones of the mirror variant (also equivalent among them in the rock-salt structure). In our data a small but clear shift is observed between the out-of-plane peak positions of the $(20)$ and $(\bar{1}1)$ rods (Fig.\ref{fig:CTR_5}-c). Similar shift is observed between $(\bar{2}0)$ and the (11) peak positions (Fig.\ref{fig:CTR_5}-d). This shift can only be taken into account if the $\vec{c}$ vector of the CoO unit cell is not perpendicular to the surface, but has an in-plane component: atoms of the fourth Co atomic layer of the oxide film are not exactly above Co atoms of the first one, characterizing a monoclinic distortion.
A second experimental observation confirms this statement. The (02), (31), and $(\bar{3}1)$ rods are equivalent in a cubic CoO(111) structure, but not in the monoclinic film. In our experiments Bragg peaks observed along the (31) and $(\bar{3}1)$ rods are wider compared to those along the (02). The (02) rod is located in the inversion-symmetry plane. Only a single component (021) Bragg peak has been measured. The wider peak, observed for the same $Q_{z}$ range along the (31) rod, corresponds to the superposition of the $(310)$ peak and the $(\bar{3}12)$ one of the mirror domain.

The monoclinic unit cell can be described using bases vector $\vec{a}=(a,0,0)$
(parallel to $\vec{a}_{Pt,1}$), $\vec{b}=(0,b,0)$ (parallel to $\vec{a}_{Pt,2}$),
and $\vec{c}=(-c_{1},c_{2},c_{3})$. As $\vec{b}$ is in the mirror
plane, if $c_{2}\neq0$ there should be 8 variants instead of the
4 experimentally observed. We have then $\vec{c}=(-c_{1},0,c_{3})$.
If we chose $\vec{c}$ as the shortest vector joining Co atoms in
the Pt (010) plane, the crystal structure is made up of a basis of
four atoms : $Co(0,0,0)$, $Co(\frac{1}{2},\frac{1}{2},0)$,$O(\frac{1}{2},0,\frac{1}{2})$,$O(0,\frac{1}{2},\frac{1}{2})$
and the reciprocal space vectors are $\vec{a^{*}}=\frac{2\pi}{a}.(1,0,\frac{c_{1}}{c_{3}})$,
$\vec{b^{*}}=\frac{2\pi}{b}.(0,1,0)$, $\vec{c^{*}}=\frac{2\pi}{c_{3}}.(0,0,1)$.
The Bragg peaks for the monoclinic system are given by $\vec{Q}(hkl)=2\pi.(\frac{h}{a},\frac{k}{b},\frac{1}{c_{3}}(l+h\frac{c_{1}}{a}))$.
We can express $\frac{c_{1}}{a}$ as $\frac{c_{1}}{a}=\frac{1}{3}+\delta$.
For the rock-salt cubic structure $\delta$=0. Thus $(200)$ and$(\overline{1}11)$
reflections have the same $Q_{z}$. For a monoclinic lattice these
two reflections are shifted by $\Delta Q=Q_{z}(2,0,0)-Q_{z}(\overline{1},1,1)=\frac{2\pi}{c_{3}}(3*\delta)$.
From Fig. \ref{fig:CTR_5}-e $\Delta Q<0$ then $\delta<0$
and $c_{1}<\frac{a}{3}$. 

Once the monoclinic distortion is qualitatively described, the structural parameters of the CoO films should be obtained. The first step is the calculations of precise in-plane parameters ${a}$ and ${b}$ from a collection of $Q_{//}$ -scans. The position of the Bragg peaks ($Q_{x}$,$Q_{y}$) were evaluated by fitting the in-plane line-shape of 12 rods belonging to different
variants. Equivalent positions were averaged to get the four inequivalent couples reported in Table \ref{TableQxQy}. Errors bars were obtained from the dispersion of the data in symmetry equivalent rods. $a$ and $b$ lattice constants were derived from the $\vec{Q}(hkl)$ formula through a weighted average of these values. Their values are ${a=5.220(2)}$ {\AA}
and ${b=3.005(1)}$ {\AA} in agreement with $h_1$ and $h_2$ values found previously.

\begin{table}[h]
\noindent \centering{}%
\begin{tabular}{|c|c|c|c|}
\hline 
rod & $Q_{x}(\mathring{A}{}^{-1})$ & $Q_{y}(\mathring{A}{}^{-1})$ & $Q_{//}(\mathring{A}{}^{-1})$\tabularnewline
\hline 
\hline 
(11) & 1.207 (2) & 2.092(1) & 2.415(2)\tabularnewline
\hline 
(20) & 2.4069(7) & 0 & 2.407(1)\tabularnewline
\hline 
(02) & 0 & 4.180(1) & 4.180(1)\tabularnewline
\hline 
(31) & 3.614(3) & 2.087(1) & 4.173(2)\tabularnewline
\hline 
\end{tabular}\caption{\label{TableQxQy} In-plane CoO peaks positions $Q_{x}$, $Q_{y}$
and $Q_{//}=\sqrt{Q_{x}^{2}+Q_{y}^{2}}$ . The small $\Delta Q_{//}$
between (11) and (20) and between (02) and (31) rod positions reveals
the in-plane distortion of the film. }
\end{table}

The intensity distribution as function of $Q_{z}$ was measured for the same 12 rods, which were treated and symmetry averaged giving the (20), (11), (31) and (02) non-equivalent rods, two of them reported in Fig.\ref{fig:CTR_5}. In the model of the CoO film with monoclinic structure, the interlayer spacing was taken uniform over the full thickness. The measured signal decreases quickly along the rods when moving farther away from the CoO Bragg peaks, which is an indication of large surface roughness expected for a polar surface. This latter was simulated by introducing an occupancy value $Occ(n)$ for each atomic layer ${n}$ of the oxide film that has been modelized by an error function erfc(z). This model is based on the assumption that the distribution of the terraces is described by a Gaussian with maximum at $z_{0}$ and with a variance ${\sigma}$. The growth mode observed here ressembles the Stranski-Krastanov mode with pyramidal islands, as observed in the case of CoO/Pt(111) over a large temperature range \cite{DeSantis2011PRB}. In this second step of the structural analysis, the interlayer spacing parameter $c_{3}$ and occupancies were obtained through a fit of the (02) rod, which is insensitive to the distortion parameter $\delta$.
After that step, occupancies are fixed and the fit of the ensemble of rods is performed to determine $c_{1}$ and optimize $c_{3}$. As ($hk)$ and ($\overline{h}k)$ rods of two mirrors variants are merged, the model structure is calculated for these two variants with identical weight.

\begin{table}[h]
\begin{centering}
\begin{tabular}{|c|c|c|c|c|}
\hline 
$a$ (\AA) & $b$ (\AA) & $c_{1}$(\AA) & $c_{3}$(\AA) & $c$ (\AA)\tabularnewline
\hline 
5.220(2) & 3.005(1) & 1.721(3) & 2.454(3) & 2.998(4)\tabularnewline
\hline 
\hline 
$\delta$ & $\beta$ (deg) & $z_{0}$ (\AA) & ${\sigma}$ (\AA) & $X^{2}$\tabularnewline
\hline 
-0.0036(7) & 125.05(7) & 43(9) & 21(4) & 2.3\tabularnewline
\hline 
\end{tabular}
\par\end{centering}

\centering{}\caption{\label{TableParam} Best-fitted parameters
of the CoO film structures. The first line concerns the basic structural
parameters. In the second line, the monoclinic distortion is highlighted
with $\delta$ and $\beta$.  $z_{0}$ and ${\sigma}$ 
are linked to the roughness of the films. ${X}^{2}$ marks
the quality of the fit.}
\end{table}

The best fit lattice parameters  are reported in Table \ref{TableParam}, as well as the values of $\delta$, $\beta=180^{\circ}-arctan(\frac{c_{3}}{c_{1}})$ and $c$.
The corresponding curves are plotted in Fig.\ref{fig:CTR_5}. In the cubic system $\beta=180^{\circ}-arctan(\sqrt{2})$=125.264${^{\circ}}$. The smaller value of $\beta=$ 125.05(7)${^{\circ}}$ expresses the deviation from cubic unit-cell.
The average film thickness is found at $4.3(9)$ nm, while ${\sigma}$ is about 1.8 nm. This should be considered only as a rough estimation, the error bar on such values being quite large. However, this estimation is in fair agreement with the thickness and roughness of the CoO layer obtained from complementary measurements (not presented here) by x-ray reflectivity ($z_{0}=3.5(6)$ and ${\sigma}$ = 0.8(3) nm) and atomic force microscopy (surface roughness rms = 1.0(2)nm).

\subsection{Implications on magnetic structure}

\begin{figure}
\resizebox{1\columnwidth}{!}{\includegraphics{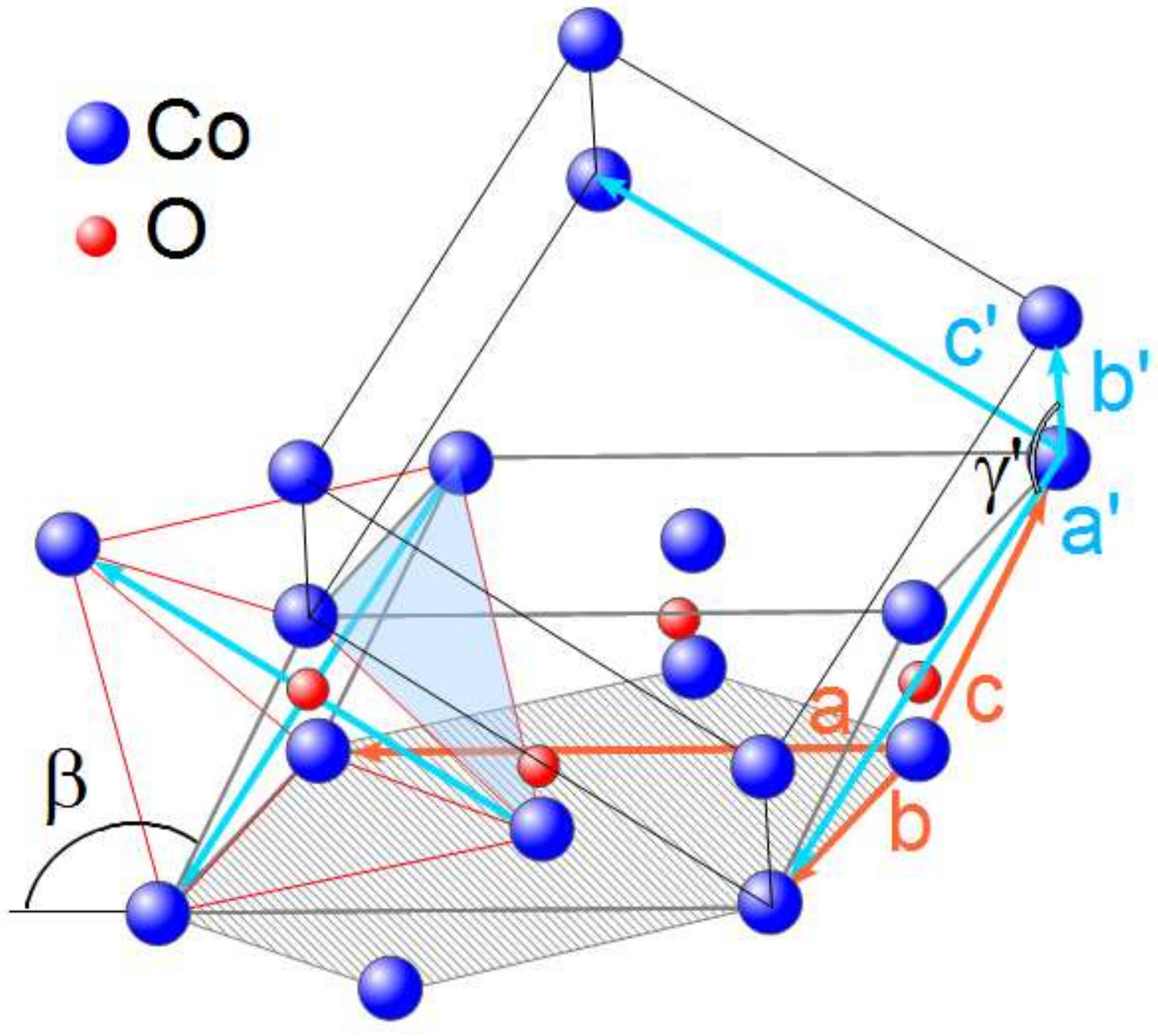}}
\caption{\label{fig:CoO_6}(color online) Illustration of CoO monoclinic ($\vec{a}$ , $\vec{b}$, and $\vec{c}$) and distorted cubic ($\vec{a'}$ , $\vec{b'}$, and $\vec{c'}$) lattice basis. ${a'}$' and ${c'}$ axes of tetragonal
elongation are indicated in the CoO${_6}$ octahedron (red lines). }
\end{figure}

\begin{table}[H]
\begin{centering}
\begin{tabular}{|c|c|c|}
\hline 
 & CoO film (this work) & CoO bulk values at 10 K\tabularnewline
 &  & (from Jauch et al.\cite{Jauch2001PRB})\tabularnewline
\hline 
$a'=b'$ (\AA) & 4.245 & 4.2682\tabularnewline
\hline 
$c'$(\AA) & 4.273 & 4.2145\tabularnewline
\hline 
$\alpha'$ (deg) & 90.00 & 90.038\tabularnewline
\hline 
$\beta'$ (deg) & 90.00 & 89.962\tabularnewline
\hline 
$\gamma'$ (deg) & 89.866 & 90.018\tabularnewline
\hline 
\hline 
$c'/a'$ & 1.007 & 0.987\tabularnewline
\hline 
$V(\mathring{A}^{3})$ & 77.00 & 76.77\tabularnewline
\hline 
$d_{Co-O}^{1}$ (\AA) & 2.137 & 2.1073\tabularnewline
\hline 
$d_{Co-O}^{2}$ (\AA) & 2.122 & 2.1341\tabularnewline
\hline 
\end{tabular}
\par\end{centering}

\centering{}\caption{\label{Tab film vs bulk-1}Comparison of CoO monoclinic phases in
a distorted unit cell. }
\end{table}

The RT monoclinic distortion imposed by epitaxy for the CoO films is of the same order of magnitude of that observed in low temperature bulk CoO phase. The structure of these two phases can be compared within the distorted cubic cell, taking as lattice basis $\vec{a'}=\vec{b}-\vec{c}$, $\vec{b'}=\vec{b}+\vec{c}$, and $\vec{c'}=\vec{a}+\vec{c}$ (Fig.\ref{fig:CoO_6}). The lattice parameters for both structures are reported in Table \ref{Tab film vs bulk-1}, along with the cell volume and the Co-O coordination distances.

In the regular cubic structure, Co atoms are embedded in the octahedral crystal field of its first neighbors O atoms. The $Co^{2+}$ ions are in the high spin $3d^{7}$ configuration, which leaves a hole in a $t_{2g}$ orbital. At local scale, the tetragonal distortion, characterised by $\frac{c'}{a'}$${\neq}1$, leads to a splitting of the $t_{2g}$ orbitals by the orthorhombic crystal field\cite{Schron_2012}. 
The elongation in the $CoO_6$ octahedron determines then the symmetry ${(x',y')}$ or ${z'}$  of the empty $t_{2g}$ orbital. The strong spin-orbit coupling completes the scenario and imposes the $Co^{2+}$ spin-orientation. 

Experiments\cite{Csiszar2005PRL} supported by theoretical calculations \cite{Boussendel2010PRB} have demonstrated such a behavior. Csiszar and coworkers\cite{Csiszar2005PRL} established the dependence of the magnitude and orientation of the magnetic moments of CoO films on the strain induced by the substrate by comparing two cases: on Ag(001), the strain induces an extensive ($\frac{c'}{a'}>1$) tetragonal deformation of the CoO layer, leading to a $Co^{2+}$ spins with  ${(x',y')}$ symmetry in the film plane. As opposite, a CoO layer sandwiched between MnO(100) layers presents a slightly compressive ($\frac{c'}{a'}<1$) deformation and have its moment directed out-of-plane, along the ${c'}$-axis.

The remarkable difference between low temperature CoO bulk and our films is the opposite $\frac{c'}{a'}$ ratio, reflecting opposite the tetragonal distortion. 
In the low temperature bulk structure, the apical Co-O interatomic distance $d_{Co-O}^{1}$ is shorter than the  $d_{Co-O}^{2}$ distance and, based only on the orbital occupation and spin-orbit coupling energy, the spin should point along the ${c'}$-axis. Nevertheless, the  spin orientation is sligthly inclined by about 11${^o}$ to 20${^o}$ from the $\vec{c'}$ axis \cite{Jauch2001PRB, Roth1958PR}. In order to minimize dipole-dipole interaction\cite{Finazzi2003PRB}, the  $Co^{2+}$ spins, which are ferromagnetically coupled, tend to be within the (111) plane of the distorted cubic cell. The competition between spin-orbit coupling and dipolar energy leads to the sligth inclination. 

In our films, the apical oxygen distance $d_{Co-O}^{1}$ is larger than the $d_{Co-O}^{2}$ distance in the ${(a',b')}$ plane and the  $t_{2g}$  hole has ${z'}$ symmetry. As a consequence, the spin should be within the ${(a',b')}$ plane, as in the case of CoO/Ag(001). 
To completely define the spin orientation, we have to consider now the dipole-dipole interaction. 
The break in cubic symmetry is identified in the film by the decrease in ${\beta}$ in the monoclinic cell. Such symmetry breaking sets the (111) plane apart, amongst the other \{111\} planes. One can then deduce that this is the plane where  $Co^{2+}$ spins are coupled ferromagnetically. To minimize the dipolar interaction\cite{Finazzi2003PRB}, the spins should be within this plane. 
As opposite to the bulk case, both interactions are not competing and may be simultaneously satisfied if the spins align along the monoclinic ${b}$-axis, i.e. parallel to the $\vec{a}_{Pt,2}$ (Fig.\ref{fig:DirectSpace} and \ref{fig:CoO_6}). 
Such a conclusion, based solely on structural distortion and symmetry arguments, is in agreement with the experimental finding by x-ray absorption spectroscopy \cite{Lamirand_PRB13}.

One may note that the RT unit cell volume of the film (77.0 $\mathring{A}^{3}$) (Table \ref{Tab film vs bulk-1}) ranges between the bulk RT rock-salt (77.36 $\mathring{A}^{3}$) and the low temperature monoclinic (76.77 $\mathring{A}^{3}$) ones. The difference is small but may be of importance for the strength of the exchange interaction. The volume decrease leads an increase in orbital superposition and consequent increase in the superexchange interaction. This might be correlated to the exceptionally high ordering temperature of the film ($T_{N}$=293 K) and to the robust perpendicular exchange bias shift, kept up to the $T_{N}$, as seen in our previous report\cite{Lamirand_PRB13}.

\section{\label{conclusion} Conclusions}

We have reported the growth and structure of ultrathin CoO layers on Pt(001) and PtFe/Pt(001) surfaces. 
The cobalt oxide grows very similarly on both surfaces with (111) rock-salt-like orientation. Its structure corresponds to a distorted hexagonal stacking with alternating planes of Co and O and displays a  rough surface. Even then, the structure develops and keeps a monoclinic distortion in the ultrathin regime. The anisotropic epitaxial mismatch with the substrates leads to a stress that deforms the Co hexagon along one of the main Pt axis; the second hexagonal layer formed by oxygen is also distorted and moves in the same direction. Such anisotropy is at the origin of the monoclinic distortion when the CoO film develops.

The magnetic performance of our sample highlights the high-quality epitaxial growth control and the importance of the strain in the stabilization of the magnetic properties.
The presence of the monoclinic phase, and the strain tensor elements of this phase, illustrates that epitaxy can be exploited to tune both the onset of the magnetic transition and the magnetic moment orientation in ultrathin films. 

\begin{acknowledgments}
Beamtime is acknowledged at the French CRG BM32/ESRF beamline. Special thanks to O. Geaymond for his invaluable technical support at the beamline. 
\end{acknowledgments}


\begin{thebibliography}{33}%
\makeatletter
\providecommand \@ifxundefined [1]{%
 \@ifx{#1\undefined}
}%
\providecommand \@ifnum [1]{%
 \ifnum #1\expandafter \@firstoftwo
 \else \expandafter \@secondoftwo
 \fi
}%
\providecommand \@ifx [1]{%
 \ifx #1\expandafter \@firstoftwo
 \else \expandafter \@secondoftwo
 \fi
}%
\providecommand \natexlab [1]{#1}%
\providecommand \enquote  [1]{``#1''}%
\providecommand \bibnamefont  [1]{#1}%
\providecommand \bibfnamefont [1]{#1}%
\providecommand \citenamefont [1]{#1}%
\providecommand \href@noop [0]{\@secondoftwo}%
\providecommand \href [0]{\begingroup \@sanitize@url \@href}%
\providecommand \@href[1]{\@@startlink{#1}\@@href}%
\providecommand \@@href[1]{\endgroup#1\@@endlink}%
\providecommand \@sanitize@url [0]{\catcode `\\12\catcode `\$12\catcode
  `\&12\catcode `\#12\catcode `\^12\catcode `\_12\catcode `\%12\relax}%
\providecommand \@@startlink[1]{}%
\providecommand \@@endlink[0]{}%
\providecommand \url  [0]{\begingroup\@sanitize@url \@url }%
\providecommand \@url [1]{\endgroup\@href {#1}{\urlprefix }}%
\providecommand \urlprefix  [0]{URL }%
\providecommand \Eprint [0]{\href }%
\providecommand \doibase [0]{http://dx.doi.org/}%
\providecommand \selectlanguage [0]{\@gobble}%
\providecommand \bibinfo  [0]{\@secondoftwo}%
\providecommand \bibfield  [0]{\@secondoftwo}%
\providecommand \translation [1]{[#1]}%
\providecommand \BibitemOpen [0]{}%
\providecommand \bibitemStop [0]{}%
\providecommand \bibitemNoStop [0]{.\EOS\space}%
\providecommand \EOS [0]{\spacefactor3000\relax}%
\providecommand \BibitemShut  [1]{\csname bibitem#1\endcsname}%
\let\auto@bib@innerbib\@empty
\bibitem [{\citenamefont {Meiklejohn}\ and\ \citenamefont
  {Bean}(1956)}]{Meiklejohn1956PR}%
  \BibitemOpen
  \bibfield  {author} {\bibinfo {author} {\bibfnamefont {W.~H.}\ \bibnamefont
  {Meiklejohn}}\ and\ \bibinfo {author} {\bibfnamefont {C.~P.}\ \bibnamefont
  {Bean}},\ }\href {\doibase 10.1103/PhysRev.102.1413} {\bibfield  {journal}
  {\bibinfo  {journal} {Phys. Rev.}\ }\textbf {\bibinfo {volume} {102}},\
  \bibinfo {pages} {1413} (\bibinfo {year} {1956})}\BibitemShut {NoStop}%
\bibitem [{\citenamefont {Nogues}\ and\ \citenamefont
  {Schuller}(1999)}]{Nogues1999}%
  \BibitemOpen
  \bibfield  {author} {\bibinfo {author} {\bibfnamefont {J.}~\bibnamefont
  {Nogues}}\ and\ \bibinfo {author} {\bibfnamefont {I.~K.}\ \bibnamefont
  {Schuller}},\ }\href@noop {} {\bibfield  {journal} {\bibinfo  {journal}
  {Journal of Magnetism and Magnetic Materials}\ }\textbf {\bibinfo {volume}
  {192}},\ \bibinfo {pages} {203 } (\bibinfo {year} {1999})}\BibitemShut
  {NoStop}%
\bibitem [{\citenamefont {Morales}\ \emph {et~al.}(2009)\citenamefont
  {Morales}, \citenamefont {Li}, \citenamefont {Olamit}, \citenamefont {Liu},
  \citenamefont {Alameda},\ and\ \citenamefont {Schuller}}]{Morales_PRB09}%
  \BibitemOpen
  \bibfield  {author} {\bibinfo {author} {\bibfnamefont {R.}~\bibnamefont
  {Morales}}, \bibinfo {author} {\bibfnamefont {Z.-P.}\ \bibnamefont {Li}},
  \bibinfo {author} {\bibfnamefont {J.}~\bibnamefont {Olamit}}, \bibinfo
  {author} {\bibfnamefont {K.}~\bibnamefont {Liu}}, \bibinfo {author}
  {\bibfnamefont {J.~M.}\ \bibnamefont {Alameda}}, \ and\ \bibinfo {author}
  {\bibfnamefont {I.~K.}\ \bibnamefont {Schuller}},\ }\href@noop {} {\bibfield
  {journal} {\bibinfo  {journal} {Phys. Rev. Lett.}\ }\textbf {\bibinfo
  {volume} {102}},\ \bibinfo {pages} {097201} (\bibinfo {year}
  {2009})}\BibitemShut {NoStop}%
\bibitem [{\citenamefont {Chappert}\ \emph {et~al.}(2007)\citenamefont
  {Chappert}, \citenamefont {Fert},\ and\ \citenamefont {Nguyen~van
  Dau}}]{Chappert2007NMat}%
  \BibitemOpen
  \bibfield  {author} {\bibinfo {author} {\bibfnamefont {C.}~\bibnamefont
  {Chappert}}, \bibinfo {author} {\bibfnamefont {A.}~\bibnamefont {Fert}}, \
  and\ \bibinfo {author} {\bibfnamefont {F.}~\bibnamefont {Nguyen~van Dau}},\
  }\href {\doibase 10.1038/nmat2024} {\bibfield  {journal} {\bibinfo  {journal}
  {Nature Materials}\ }\textbf {\bibinfo {volume} {6}},\ \bibinfo {pages} {813}
  (\bibinfo {year} {2007})}\BibitemShut {NoStop}%
\bibitem [{\citenamefont {Kiwi}(2001)}]{Kiwi_JMMM2001}%
  \BibitemOpen
  \bibfield  {author} {\bibinfo {author} {\bibfnamefont {M.}~\bibnamefont
  {Kiwi}},\ }\href@noop {} {\bibfield  {journal} {\bibinfo  {journal} {J. Magn.
  Magn. Mater.}\ }\textbf {\bibinfo {volume} {234}},\ \bibinfo {pages} {584}
  (\bibinfo {year} {2001})}\BibitemShut {NoStop}%
\bibitem [{\citenamefont {Ambrose}\ and\ \citenamefont
  {Chien}(1996)}]{Ambrose_1996}%
  \BibitemOpen
  \bibfield  {author} {\bibinfo {author} {\bibfnamefont {T.}~\bibnamefont
  {Ambrose}}\ and\ \bibinfo {author} {\bibfnamefont {C.~L.}\ \bibnamefont
  {Chien}},\ }\href {\doibase 10.1103/PhysRevLett.76.1743} {\bibfield
  {journal} {\bibinfo  {journal} {Phys. Rev. Lett.}\ }\textbf {\bibinfo
  {volume} {76}},\ \bibinfo {pages} {1743} (\bibinfo {year}
  {1996})}\BibitemShut {NoStop}%
\bibitem [{\citenamefont {Abarra}\ \emph {et~al.}(1996)\citenamefont {Abarra},
  \citenamefont {Takano}, \citenamefont {Hellman},\ and\ \citenamefont
  {Berkowitz}}]{Abarra_1996}%
  \BibitemOpen
  \bibfield  {author} {\bibinfo {author} {\bibfnamefont {E.~N.}\ \bibnamefont
  {Abarra}}, \bibinfo {author} {\bibfnamefont {K.}~\bibnamefont {Takano}},
  \bibinfo {author} {\bibfnamefont {F.}~\bibnamefont {Hellman}}, \ and\
  \bibinfo {author} {\bibfnamefont {A.~E.}\ \bibnamefont {Berkowitz}},\ }\href
  {\doibase 10.1103/PhysRevLett.77.3451} {\bibfield  {journal} {\bibinfo
  {journal} {Phys. Rev. Lett.}\ }\textbf {\bibinfo {volume} {77}},\ \bibinfo
  {pages} {3451} (\bibinfo {year} {1996})}\BibitemShut {NoStop}%
\bibitem [{\citenamefont {Tang}\ \emph {et~al.}(2003)\citenamefont {Tang},
  \citenamefont {Smith}, \citenamefont {Zink}, \citenamefont {Hellman},\ and\
  \citenamefont {Berkowitz}}]{Tang_2003}%
  \BibitemOpen
  \bibfield  {author} {\bibinfo {author} {\bibfnamefont {Y.~J.}\ \bibnamefont
  {Tang}}, \bibinfo {author} {\bibfnamefont {D.~J.}\ \bibnamefont {Smith}},
  \bibinfo {author} {\bibfnamefont {B.~L.}\ \bibnamefont {Zink}}, \bibinfo
  {author} {\bibfnamefont {F.}~\bibnamefont {Hellman}}, \ and\ \bibinfo
  {author} {\bibfnamefont {A.~E.}\ \bibnamefont {Berkowitz}},\ }\href@noop {}
  {\bibfield  {journal} {\bibinfo  {journal} {Phys. Rev. B}\ }\textbf {\bibinfo
  {volume} {67}},\ \bibinfo {pages} {054408} (\bibinfo {year}
  {2003})}\BibitemShut {NoStop}%
\bibitem [{\citenamefont {Jauch}\ \emph {et~al.}(2001)\citenamefont {Jauch},
  \citenamefont {Reehuis}, \citenamefont {Bleif}, \citenamefont {Kubanek},\
  and\ \citenamefont {Pattison}}]{Jauch2001PRB}%
  \BibitemOpen
  \bibfield  {author} {\bibinfo {author} {\bibfnamefont {W.}~\bibnamefont
  {Jauch}}, \bibinfo {author} {\bibfnamefont {M.}~\bibnamefont {Reehuis}},
  \bibinfo {author} {\bibfnamefont {H.~J.}\ \bibnamefont {Bleif}}, \bibinfo
  {author} {\bibfnamefont {F.}~\bibnamefont {Kubanek}}, \ and\ \bibinfo
  {author} {\bibfnamefont {P.}~\bibnamefont {Pattison}},\ }\href {\doibase
  10.1103/PhysRevB.64.052102} {\bibfield  {journal} {\bibinfo  {journal} {Phys.
  Rev. B}\ }\textbf {\bibinfo {volume} {64}},\ \bibinfo {pages} {052102}
  (\bibinfo {year} {2001})}\BibitemShut {NoStop}%
\bibitem [{\citenamefont {Roth}(1958)}]{Roth1958PR}%
  \BibitemOpen
  \bibfield  {author} {\bibinfo {author} {\bibfnamefont {W.~L.}\ \bibnamefont
  {Roth}},\ }\href@noop {} {\bibfield  {journal} {\bibinfo  {journal} {Phys.\
  Rev.}\ }\textbf {\bibinfo {volume} {110}},\ \bibinfo {pages} {1333} (\bibinfo
  {year} {1958})}\BibitemShut {NoStop}%
\bibitem [{\citenamefont {Anderson}(1950)}]{Anderson_1950}%
  \BibitemOpen
  \bibfield  {author} {\bibinfo {author} {\bibfnamefont {P.~W.}\ \bibnamefont
  {Anderson}},\ }\href {\doibase 10.1103/PhysRev.79.350} {\bibfield  {journal}
  {\bibinfo  {journal} {Phys. Rev.}\ }\textbf {\bibinfo {volume} {79}},\
  \bibinfo {pages} {350} (\bibinfo {year} {1950})}\BibitemShut {NoStop}%
\bibitem [{\citenamefont {Anderson}(1963)}]{Anderson_1963}%
  \BibitemOpen
  \bibfield  {author} {\bibinfo {author} {\bibfnamefont {P.~W.}\ \bibnamefont
  {Anderson}},\ }\href@noop {} {\emph {\bibinfo {title} {Theory of Magnetic
  Exchange Interactions: Exchange in Insulators and Semiconductors}}},\ edited
  by\ \bibinfo {editor} {\bibfnamefont {F.}~\bibnamefont {Seitz}}\ and\
  \bibinfo {editor} {\bibfnamefont {D.}~\bibnamefont {Turnbull}},\ \bibinfo
  {series} {Solid State Physics}, Vol.~\bibinfo {volume} {14}\ (\bibinfo
  {publisher} {Academic Press},\ \bibinfo {year} {1963})\ pp.\ \bibinfo {pages}
  {99 --214}\BibitemShut {NoStop}%
\bibitem [{\citenamefont {Schindler}\ \emph {et~al.}(2009)\citenamefont
  {Schindler}, \citenamefont {Wang}, \citenamefont {Chass\'{e}}, \citenamefont
  {Neddermeyer},\ and\ \citenamefont {Widdra}}]{Schindler2009}%
  \BibitemOpen
  \bibfield  {author} {\bibinfo {author} {\bibfnamefont {K.-M.}\ \bibnamefont
  {Schindler}}, \bibinfo {author} {\bibfnamefont {J.}~\bibnamefont {Wang}},
  \bibinfo {author} {\bibfnamefont {A.}~\bibnamefont {Chass\'{e}}}, \bibinfo
  {author} {\bibfnamefont {H.}~\bibnamefont {Neddermeyer}}, \ and\ \bibinfo
  {author} {\bibfnamefont {W.}~\bibnamefont {Widdra}},\ }\href {\doibase
  10.1016/j.susc.2009.06.020} {\bibfield  {journal} {\bibinfo  {journal}
  {Surface Science}\ }\textbf {\bibinfo {volume} {603}},\ \bibinfo {pages}
  {2658} (\bibinfo {year} {2009})}\BibitemShut {NoStop}%
\bibitem [{\citenamefont {Gragnaniello}\ \emph {et~al.}(2010)\citenamefont
  {Gragnaniello}, \citenamefont {Agnoli}, \citenamefont {Parteder},
  \citenamefont {Barolo}, \citenamefont {Bondino}, \citenamefont {Allegretti},
  \citenamefont {Surnev}, \citenamefont {Granozzi},\ and\ \citenamefont
  {Netzer}}]{Gragnaniello2010a}%
  \BibitemOpen
  \bibfield  {author} {\bibinfo {author} {\bibfnamefont {L.}~\bibnamefont
  {Gragnaniello}}, \bibinfo {author} {\bibfnamefont {S.}~\bibnamefont
  {Agnoli}}, \bibinfo {author} {\bibfnamefont {G.}~\bibnamefont {Parteder}},
  \bibinfo {author} {\bibfnamefont {A.}~\bibnamefont {Barolo}}, \bibinfo
  {author} {\bibfnamefont {F.}~\bibnamefont {Bondino}}, \bibinfo {author}
  {\bibfnamefont {F.}~\bibnamefont {Allegretti}}, \bibinfo {author}
  {\bibfnamefont {S.}~\bibnamefont {Surnev}}, \bibinfo {author} {\bibfnamefont
  {G.}~\bibnamefont {Granozzi}}, \ and\ \bibinfo {author} {\bibfnamefont
  {F.~P.}\ \bibnamefont {Netzer}},\ }\href {\doibase
  10.1016/j.susc.2010.08.012} {\bibfield  {journal} {\bibinfo  {journal}
  {Surface Science}\ }\textbf {\bibinfo {volume} {604}},\ \bibinfo {pages}
  {2002} (\bibinfo {year} {2010})}\BibitemShut {NoStop}%
\bibitem [{\citenamefont {Meyer}\ \emph {et~al.}(2008)\citenamefont {Meyer},
  \citenamefont {Hock}, \citenamefont {Biedermann}, \citenamefont {Gubo},
  \citenamefont {M\"uller}, \citenamefont {Hammer},\ and\ \citenamefont
  {Heinz}}]{Meyer_PRL08}%
  \BibitemOpen
  \bibfield  {author} {\bibinfo {author} {\bibfnamefont {W.}~\bibnamefont
  {Meyer}}, \bibinfo {author} {\bibfnamefont {D.}~\bibnamefont {Hock}},
  \bibinfo {author} {\bibfnamefont {K.}~\bibnamefont {Biedermann}}, \bibinfo
  {author} {\bibfnamefont {M.}~\bibnamefont {Gubo}}, \bibinfo {author}
  {\bibfnamefont {S.}~\bibnamefont {M\"uller}}, \bibinfo {author}
  {\bibfnamefont {L.}~\bibnamefont {Hammer}}, \ and\ \bibinfo {author}
  {\bibfnamefont {K.}~\bibnamefont {Heinz}},\ }\href {\doibase
  10.1103/PhysRevLett.101.016103} {\bibfield  {journal} {\bibinfo  {journal}
  {Phys. Rev. Lett.}\ }\textbf {\bibinfo {volume} {101}},\ \bibinfo {pages}
  {016103} (\bibinfo {year} {2008})}\BibitemShut {NoStop}%
\bibitem [{\citenamefont {Gubo}\ \emph {et~al.}(2012)\citenamefont {Gubo},
  \citenamefont {Ebensperger}, \citenamefont {Meyer}, \citenamefont {Hammer},
  \citenamefont {Heinz}, \citenamefont {Mittendorfer},\ and\ \citenamefont
  {Redinger}}]{Gubo_PRL12}%
  \BibitemOpen
  \bibfield  {author} {\bibinfo {author} {\bibfnamefont {M.}~\bibnamefont
  {Gubo}}, \bibinfo {author} {\bibfnamefont {C.}~\bibnamefont {Ebensperger}},
  \bibinfo {author} {\bibfnamefont {W.}~\bibnamefont {Meyer}}, \bibinfo
  {author} {\bibfnamefont {L.}~\bibnamefont {Hammer}}, \bibinfo {author}
  {\bibfnamefont {K.}~\bibnamefont {Heinz}}, \bibinfo {author} {\bibfnamefont
  {F.}~\bibnamefont {Mittendorfer}}, \ and\ \bibinfo {author} {\bibfnamefont
  {J.}~\bibnamefont {Redinger}},\ }\href {\doibase
  10.1103/PhysRevLett.108.066101} {\bibfield  {journal} {\bibinfo  {journal}
  {Phys. Rev. Lett.}\ }\textbf {\bibinfo {volume} {108}},\ \bibinfo {pages}
  {066101} (\bibinfo {year} {2012})}\BibitemShut {NoStop}%
\bibitem [{\citenamefont {De~Santis}\ \emph {et~al.}(2011)\citenamefont
  {De~Santis}, \citenamefont {Buchsbaum}, \citenamefont {Varga},\ and\
  \citenamefont {Schmid}}]{DeSantis2011PRB}%
  \BibitemOpen
  \bibfield  {author} {\bibinfo {author} {\bibfnamefont {M.}~\bibnamefont
  {De~Santis}}, \bibinfo {author} {\bibfnamefont {A.}~\bibnamefont
  {Buchsbaum}}, \bibinfo {author} {\bibfnamefont {P.}~\bibnamefont {Varga}}, \
  and\ \bibinfo {author} {\bibfnamefont {M.}~\bibnamefont {Schmid}},\ }\href
  {\doibase 10.1103/PhysRevB.84.125430} {\bibfield  {journal} {\bibinfo
  {journal} {Phys. Rev. B}\ }\textbf {\bibinfo {volume} {84}},\ \bibinfo
  {pages} {125430} (\bibinfo {year} {2011})}\BibitemShut {NoStop}%
\bibitem [{\citenamefont {Csiszar}\ \emph {et~al.}(2005)\citenamefont
  {Csiszar}, \citenamefont {Haverkort}, \citenamefont {Hu}, \citenamefont
  {Tanaka}, \citenamefont {Hsieh}, \citenamefont {Lin}, \citenamefont {Chen},
  \citenamefont {Hibma},\ and\ \citenamefont {Tjeng}}]{Csiszar2005PRL}%
  \BibitemOpen
  \bibfield  {author} {\bibinfo {author} {\bibfnamefont {S.~I.}\ \bibnamefont
  {Csiszar}}, \bibinfo {author} {\bibfnamefont {M.~W.}\ \bibnamefont
  {Haverkort}}, \bibinfo {author} {\bibfnamefont {Z.}~\bibnamefont {Hu}},
  \bibinfo {author} {\bibfnamefont {A.}~\bibnamefont {Tanaka}}, \bibinfo
  {author} {\bibfnamefont {H.~H.}\ \bibnamefont {Hsieh}}, \bibinfo {author}
  {\bibfnamefont {H.-J.}\ \bibnamefont {Lin}}, \bibinfo {author} {\bibfnamefont
  {C.~T.}\ \bibnamefont {Chen}}, \bibinfo {author} {\bibfnamefont
  {T.}~\bibnamefont {Hibma}}, \ and\ \bibinfo {author} {\bibfnamefont {L.~H.}\
  \bibnamefont {Tjeng}},\ }\href {\doibase 10.1103/PhysRevLett.95.187205}
  {\bibfield  {journal} {\bibinfo  {journal} {Phys. Rev. Lett.}\ }\textbf
  {\bibinfo {volume} {95}},\ \bibinfo {pages} {187205} (\bibinfo {year}
  {2005})}\BibitemShut {NoStop}%
\bibitem [{\citenamefont {Baudoing-Savois}\ \emph {et~al.}(1999)\citenamefont
  {Baudoing-Savois}, \citenamefont {Santis}, \citenamefont {Saint-Lager},
  \citenamefont {Dolle}, \citenamefont {Geaymond}, \citenamefont {Taunier},
  \citenamefont {Jeantet}, \citenamefont {Roux}, \citenamefont {Renaud},
  \citenamefont {Barbier}, \citenamefont {Robach}, \citenamefont {Ulrich},
  \citenamefont {Mougin},\ and\ \citenamefont
  {B\'{e}rard}}]{BaudoingSavois_1999_213}%
  \BibitemOpen
  \bibfield  {author} {\bibinfo {author} {\bibfnamefont {R.}~\bibnamefont
  {Baudoing-Savois}}, \bibinfo {author} {\bibfnamefont {M.~D.}\ \bibnamefont
  {Santis}}, \bibinfo {author} {\bibfnamefont {M.}~\bibnamefont {Saint-Lager}},
  \bibinfo {author} {\bibfnamefont {P.}~\bibnamefont {Dolle}}, \bibinfo
  {author} {\bibfnamefont {O.}~\bibnamefont {Geaymond}}, \bibinfo {author}
  {\bibfnamefont {P.}~\bibnamefont {Taunier}}, \bibinfo {author} {\bibfnamefont
  {P.}~\bibnamefont {Jeantet}}, \bibinfo {author} {\bibfnamefont
  {J.}~\bibnamefont {Roux}}, \bibinfo {author} {\bibfnamefont {G.}~\bibnamefont
  {Renaud}}, \bibinfo {author} {\bibfnamefont {A.}~\bibnamefont {Barbier}},
  \bibinfo {author} {\bibfnamefont {O.}~\bibnamefont {Robach}}, \bibinfo
  {author} {\bibfnamefont {O.}~\bibnamefont {Ulrich}}, \bibinfo {author}
  {\bibfnamefont {A.}~\bibnamefont {Mougin}}, \ and\ \bibinfo {author}
  {\bibfnamefont {G.}~\bibnamefont {B\'{e}rard}},\ }\href@noop {} {\bibfield
  {journal} {\bibinfo  {journal} {Nucl. Instrum. Meth. B}\ }\textbf {\bibinfo
  {volume} {149}},\ \bibinfo {pages} {213} (\bibinfo {year}
  {1999})}\BibitemShut {NoStop}%
\bibitem [{\citenamefont {Abernathy}\ \emph {et~al.}(1992)\citenamefont
  {Abernathy}, \citenamefont {Mochrie}, \citenamefont {Zehner}, \citenamefont
  {Grubel},\ and\ \citenamefont {Gibbs}}]{AbernathyDL_PRB1992}%
  \BibitemOpen
  \bibfield  {author} {\bibinfo {author} {\bibfnamefont {D.~L.}\ \bibnamefont
  {Abernathy}}, \bibinfo {author} {\bibfnamefont {S.~G.~J.}\ \bibnamefont
  {Mochrie}}, \bibinfo {author} {\bibfnamefont {D.~M.}\ \bibnamefont {Zehner}},
  \bibinfo {author} {\bibfnamefont {G.}~\bibnamefont {Grubel}}, \ and\ \bibinfo
  {author} {\bibfnamefont {D.}~\bibnamefont {Gibbs}},\ }\href@noop {}
  {\bibfield  {journal} {\bibinfo  {journal} {Phys. Rev. B}\ }\textbf {\bibinfo
  {volume} {45}},\ \bibinfo {pages} {9272} (\bibinfo {year}
  {1992})}\BibitemShut {NoStop}%
\bibitem [{\citenamefont {Soares}\ \emph {et~al.}(2012)\citenamefont {Soares},
  \citenamefont {De~Santis}, \citenamefont {Tolentino}, \citenamefont {Ramos},
  \citenamefont {El~Jawad}, \citenamefont {Gauthier}, \citenamefont {Yildiz},\
  and\ \citenamefont {Przybylski}}]{Soares2012PRB}%
  \BibitemOpen
  \bibfield  {author} {\bibinfo {author} {\bibfnamefont {M.~M.}\ \bibnamefont
  {Soares}}, \bibinfo {author} {\bibfnamefont {M.}~\bibnamefont {De~Santis}},
  \bibinfo {author} {\bibfnamefont {H.~C.~N.}\ \bibnamefont {Tolentino}},
  \bibinfo {author} {\bibfnamefont {A.~Y.}\ \bibnamefont {Ramos}}, \bibinfo
  {author} {\bibfnamefont {M.}~\bibnamefont {El~Jawad}}, \bibinfo {author}
  {\bibfnamefont {Y.}~\bibnamefont {Gauthier}}, \bibinfo {author}
  {\bibfnamefont {F.}~\bibnamefont {Yildiz}}, \ and\ \bibinfo {author}
  {\bibfnamefont {M.}~\bibnamefont {Przybylski}},\ }\href@noop {} {\bibfield
  {journal} {\bibinfo  {journal} {Phys. Rev. B}\ }\textbf {\bibinfo {volume}
  {85}},\ \bibinfo {pages} {205417} (\bibinfo {year} {2012})}\BibitemShut
  {NoStop}%
\bibitem [{\citenamefont {Robinson}(1986)}]{RobinsonIK_PRB1986}%
  \BibitemOpen
  \bibfield  {author} {\bibinfo {author} {\bibfnamefont {I.~K.}\ \bibnamefont
  {Robinson}},\ }\href {\doibase 10.1103/PhysRevB.33.3830} {\bibfield
  {journal} {\bibinfo  {journal} {Phys. Rev. B}\ }\textbf {\bibinfo {volume}
  {33}},\ \bibinfo {pages} {3830} (\bibinfo {year} {1986})}\BibitemShut
  {NoStop}%
\bibitem [{\citenamefont {Vlieg}(2000)}]{VliegE_JACrys2000:ROD}%
  \BibitemOpen
  \bibfield  {author} {\bibinfo {author} {\bibfnamefont {E.}~\bibnamefont
  {Vlieg}},\ }\href@noop {} {\bibfield  {journal} {\bibinfo  {journal} {J.
  Appl. Crystallogr.}\ }\textbf {\bibinfo {volume} {33}},\ \bibinfo {pages}
  {401} (\bibinfo {year} {2000})}\BibitemShut {NoStop}%
\bibitem [{\citenamefont {Vlieg}(1997)}]{VliegE_JACryst1997}%
  \BibitemOpen
  \bibfield  {author} {\bibinfo {author} {\bibfnamefont {E.}~\bibnamefont
  {Vlieg}},\ }\href@noop {} {\bibfield  {journal} {\bibinfo  {journal} {J.
  Appl. Crystallogr.}\ }\textbf {\bibinfo {volume} {30}},\ \bibinfo {pages}
  {532} (\bibinfo {year} {1997})}\BibitemShut {NoStop}%
\bibitem [{\citenamefont {He}\ \emph {et~al.}(2005)\citenamefont {He},
  \citenamefont {Zhang}, \citenamefont {Ma}, \citenamefont {Jia}, \citenamefont
  {Xue},\ and\ \citenamefont {Qiu}}]{He_PRB2005}%
  \BibitemOpen
  \bibfield  {author} {\bibinfo {author} {\bibfnamefont {K.}~\bibnamefont
  {He}}, \bibinfo {author} {\bibfnamefont {L.~J.}\ \bibnamefont {Zhang}},
  \bibinfo {author} {\bibfnamefont {X.~C.}\ \bibnamefont {Ma}}, \bibinfo
  {author} {\bibfnamefont {J.~F.}\ \bibnamefont {Jia}}, \bibinfo {author}
  {\bibfnamefont {Q.~K.}\ \bibnamefont {Xue}}, \ and\ \bibinfo {author}
  {\bibfnamefont {Z.~Q.}\ \bibnamefont {Qiu}},\ }\href@noop {} {\bibfield
  {journal} {\bibinfo  {journal} {Phys. Rev. B}\ }\textbf {\bibinfo {volume}
  {72}},\ \bibinfo {pages} {155432} (\bibinfo {year} {2005})}\BibitemShut
  {NoStop}%
\bibitem [{\citenamefont {Soares}\ \emph {et~al.}(2011)\citenamefont {Soares},
  \citenamefont {Tolentino}, \citenamefont {Santis}, \citenamefont {Ramos},\
  and\ \citenamefont {Cezar}}]{Soares2011JAP}%
  \BibitemOpen
  \bibfield  {author} {\bibinfo {author} {\bibfnamefont {M.~M.}\ \bibnamefont
  {Soares}}, \bibinfo {author} {\bibfnamefont {H.~C.~N.}\ \bibnamefont
  {Tolentino}}, \bibinfo {author} {\bibfnamefont {M.~D.}\ \bibnamefont
  {Santis}}, \bibinfo {author} {\bibfnamefont {A.~Y.}\ \bibnamefont {Ramos}}, \
  and\ \bibinfo {author} {\bibfnamefont {J.~C.}\ \bibnamefont {Cezar}},\
  }\href@noop {} {\bibfield  {journal} {\bibinfo  {journal} {J. Appl. Phys.}\
  }\textbf {\bibinfo {volume} {109}},\ \bibinfo {eid} {07D725} (\bibinfo {year}
  {2011})}\BibitemShut {NoStop}%
\bibitem [{\citenamefont {Warren}(1969)}]{Warren_1969}%
  \BibitemOpen
  \bibfield  {author} {\bibinfo {author} {\bibfnamefont {B.}~\bibnamefont
  {Warren}},\ }\href@noop {} {\emph {\bibinfo {title} {X-ray diffraction}}}\
  (\bibinfo  {publisher} {Addison-Wesley, Reading},\ \bibinfo {year} {1969})\
  p.\ \bibinfo {pages} {382}\BibitemShut {NoStop}%
\bibitem [{\citenamefont {Regan}\ \emph {et~al.}(2001)\citenamefont {Regan},
  \citenamefont {Ohldag}, \citenamefont {Stamm}, \citenamefont {Nolting},
  \citenamefont {L\"uning}, \citenamefont {St\"ohr},\ and\ \citenamefont
  {White}}]{Regan2001PRB}%
  \BibitemOpen
  \bibfield  {author} {\bibinfo {author} {\bibfnamefont {T.~J.}\ \bibnamefont
  {Regan}}, \bibinfo {author} {\bibfnamefont {H.}~\bibnamefont {Ohldag}},
  \bibinfo {author} {\bibfnamefont {C.}~\bibnamefont {Stamm}}, \bibinfo
  {author} {\bibfnamefont {F.}~\bibnamefont {Nolting}}, \bibinfo {author}
  {\bibfnamefont {J.}~\bibnamefont {L\"uning}}, \bibinfo {author}
  {\bibfnamefont {J.}~\bibnamefont {St\"ohr}}, \ and\ \bibinfo {author}
  {\bibfnamefont {R.~L.}\ \bibnamefont {White}},\ }\href {\doibase
  10.1103/PhysRevB.64.214422} {\bibfield  {journal} {\bibinfo  {journal} {Phys.
  Rev. B}\ }\textbf {\bibinfo {volume} {64}},\ \bibinfo {pages} {214422}
  (\bibinfo {year} {2001})}\BibitemShut {NoStop}%
\bibitem [{\citenamefont {Bali}\ \emph {et~al.}(2012)\citenamefont {Bali},
  \citenamefont {Soares}, \citenamefont {Ramos}, \citenamefont {Tolentino},
  \citenamefont {Yildiz}, \citenamefont {Boudot}, \citenamefont {Proux},
  \citenamefont {De~Santis}, \citenamefont {Przybylski},\ and\ \citenamefont
  {Kirschner}}]{Bali_APL12}%
  \BibitemOpen
  \bibfield  {author} {\bibinfo {author} {\bibfnamefont {R.}~\bibnamefont
  {Bali}}, \bibinfo {author} {\bibfnamefont {M.~M.}\ \bibnamefont {Soares}},
  \bibinfo {author} {\bibfnamefont {A.~Y.}\ \bibnamefont {Ramos}}, \bibinfo
  {author} {\bibfnamefont {H.~C.~N.}\ \bibnamefont {Tolentino}}, \bibinfo
  {author} {\bibfnamefont {F.}~\bibnamefont {Yildiz}}, \bibinfo {author}
  {\bibfnamefont {C.}~\bibnamefont {Boudot}}, \bibinfo {author} {\bibfnamefont
  {O.}~\bibnamefont {Proux}}, \bibinfo {author} {\bibfnamefont
  {M.}~\bibnamefont {De~Santis}}, \bibinfo {author} {\bibfnamefont
  {M.}~\bibnamefont {Przybylski}}, \ and\ \bibinfo {author} {\bibfnamefont
  {J.}~\bibnamefont {Kirschner}},\ }\href@noop {} {\bibfield  {journal}
  {\bibinfo  {journal} {Appl. Phys. Lett.}\ }\textbf {\bibinfo {volume}
  {{100}}},\ \bibinfo {pages} {{132403}} (\bibinfo {year}
  {{2012}})}\BibitemShut {NoStop}%
\bibitem [{\citenamefont {Lamirand}\ \emph {et~al.}(2013)\citenamefont
  {Lamirand}, \citenamefont {Soares}, \citenamefont {Ramos}, \citenamefont
  {Tolentino}, \citenamefont {{De Santis}}, \citenamefont {Cezar},
  \citenamefont {de~Siervo},\ and\ \citenamefont {Jamet}}]{Lamirand_PRB13}%
  \BibitemOpen
  \bibfield  {author} {\bibinfo {author} {\bibfnamefont {A.~D.}\ \bibnamefont
  {Lamirand}}, \bibinfo {author} {\bibfnamefont {M.~M.}\ \bibnamefont
  {Soares}}, \bibinfo {author} {\bibfnamefont {A.~Y.}\ \bibnamefont {Ramos}},
  \bibinfo {author} {\bibfnamefont {H.~C.~N.}\ \bibnamefont {Tolentino}},
  \bibinfo {author} {\bibfnamefont {M.}~\bibnamefont {{De Santis}}}, \bibinfo
  {author} {\bibfnamefont {J.~C.}\ \bibnamefont {Cezar}}, \bibinfo {author}
  {\bibfnamefont {A.}~\bibnamefont {de~Siervo}}, \ and\ \bibinfo {author}
  {\bibfnamefont {M.}~\bibnamefont {Jamet}},\ }\href@noop {} {\bibfield
  {journal} {\bibinfo  {journal} {Phys. Rev. B}\ }\textbf {\bibinfo {volume}
  {88}},\ \bibinfo {pages} {140401} (\bibinfo {year} {2013})}\BibitemShut
  {NoStop}%
\bibitem [{\citenamefont {Schr\"on}\ \emph {et~al.}(2012)\citenamefont
  {Schr\"on}, \citenamefont {R\"odl},\ and\ \citenamefont
  {Bechstedt}}]{Schron_2012}%
  \BibitemOpen
  \bibfield  {author} {\bibinfo {author} {\bibfnamefont {A.}~\bibnamefont
  {Schr\"on}}, \bibinfo {author} {\bibfnamefont {C.}~\bibnamefont {R\"odl}}, \
  and\ \bibinfo {author} {\bibfnamefont {F.}~\bibnamefont {Bechstedt}},\ }\href
  {\doibase 10.1103/PhysRevB.86.115134} {\bibfield  {journal} {\bibinfo
  {journal} {Phys. Rev. B}\ }\textbf {\bibinfo {volume} {86}},\ \bibinfo
  {pages} {115134} (\bibinfo {year} {2012})}\BibitemShut {NoStop}%
\bibitem [{\citenamefont {Boussendel}\ \emph {et~al.}(2010)\citenamefont
  {Boussendel}, \citenamefont {Baadji}, \citenamefont {Haroun}, \citenamefont
  {Dreyss\'e},\ and\ \citenamefont {Alouani}}]{Boussendel2010PRB}%
  \BibitemOpen
  \bibfield  {author} {\bibinfo {author} {\bibfnamefont {A.}~\bibnamefont
  {Boussendel}}, \bibinfo {author} {\bibfnamefont {N.}~\bibnamefont {Baadji}},
  \bibinfo {author} {\bibfnamefont {A.}~\bibnamefont {Haroun}}, \bibinfo
  {author} {\bibfnamefont {H.}~\bibnamefont {Dreyss\'e}}, \ and\ \bibinfo
  {author} {\bibfnamefont {M.}~\bibnamefont {Alouani}},\ }\href {\doibase
  10.1103/PhysRevB.81.184432} {\bibfield  {journal} {\bibinfo  {journal} {Phys.
  Rev. B}\ }\textbf {\bibinfo {volume} {81}},\ \bibinfo {pages} {184432}
  (\bibinfo {year} {2010})}\BibitemShut {NoStop}%
\bibitem [{\citenamefont {Finazzi}\ and\ \citenamefont
  {Altieri}(2003)}]{Finazzi2003PRB}%
  \BibitemOpen
  \bibfield  {author} {\bibinfo {author} {\bibfnamefont {M.}~\bibnamefont
  {Finazzi}}\ and\ \bibinfo {author} {\bibfnamefont {S.}~\bibnamefont
  {Altieri}},\ }\href {\doibase 10.1103/PhysRevB.68.054420} {\bibfield
  {journal} {\bibinfo  {journal} {Phys. Rev. B}\ }\textbf {\bibinfo {volume}
  {68}},\ \bibinfo {pages} {054420} (\bibinfo {year} {2003})}\BibitemShut
  {NoStop}%
\end{thebibliography}
\end{document}